\documentclass[%
 reprint,
superscriptaddress,
nofootinbib,
 amsmath,amssymb,
aps,
pre,
]{revtex4-2}

\usepackage{color}
\usepackage[normalem]{ulem}
\usepackage{placeins}

\usepackage{graphicx}% Include figure files
\usepackage{dcolumn}% Align table columns on decimal point
\usepackage{bm}% bold math

\usepackage{rotating}
\usepackage{multirow}

\usepackage{booktabs}% for \midrule and \cmidrule macros
\newcommand\headercell[1]{%
   \smash[b]{\begin{tabular}[t]{@{}c@{}} #1 \end{tabular}}}

\begin{document}

\preprint{APS/123-QED}

\title{Studying First Passage Problems using Neural Networks: A Case Study in the Slit-Well Microfluidic Device}% Force line breaks with \\

\author{Andrew Nagel}
\affiliation{%
 Faculty of Science,
 University of Ontario Institute of Technology,
 2000 Simcoe St N, 
 Oshawa, Ontario, Canada L1H7K4
 }
\author{Martin Magill}
\affiliation{%
 Faculty of Science,
 University of Ontario Institute of Technology,
 2000 Simcoe St N, 
 Oshawa, Ontario, Canada L1H7K4
 }
\author{Hendrick W. de Haan}%
 \email{Hendrick.deHaan@uoit.ca}
\affiliation{%
 Faculty of Science,
 University of Ontario Institute of Technology,
 2000 Simcoe St N, 
 Oshawa, Ontario, Canada L1H7K4
 }%

\date{\today}% It is always \today, today,

\widetext
\begin{abstract}

This study presents deep neural network solutions to a time-integrated Smoluchowski equation modeling the mean first passage time of nanoparticles traversing the slit-well microfluidic device.
This physical scenario is representative of a broader class of parameterized first passage problems in which key output metrics are dictated by a complicated interplay of problem parameters and system geometry.
Specifically, whereas these types of problems are commonly studied using particle simulations of stochastic differential equation models, here the corresponding partial differential equation model is solved using a method based on deep neural networks.
The results illustrate that the neural network method is synergistic with the time-integrated Smoluchowski model: 
together, these are used to construct continuous mappings from key physical inputs (applied voltage and particle diameter) to key output metrics (mean first passage time and effective mobility).
In particular, this capability is a unique advantage of the time-integrated Smoluchowski model over the corresponding stochastic differential equation models.
Furthermore, the neural network method is demonstrated to easily and reliably handle geometry-modifying parameters, which is generally difficult to accomplish using other methods.

\end{abstract}

\maketitle

\section{Introduction \label{sec:intro}}

\begin{figure}
\includegraphics[width=0.99\columnwidth]{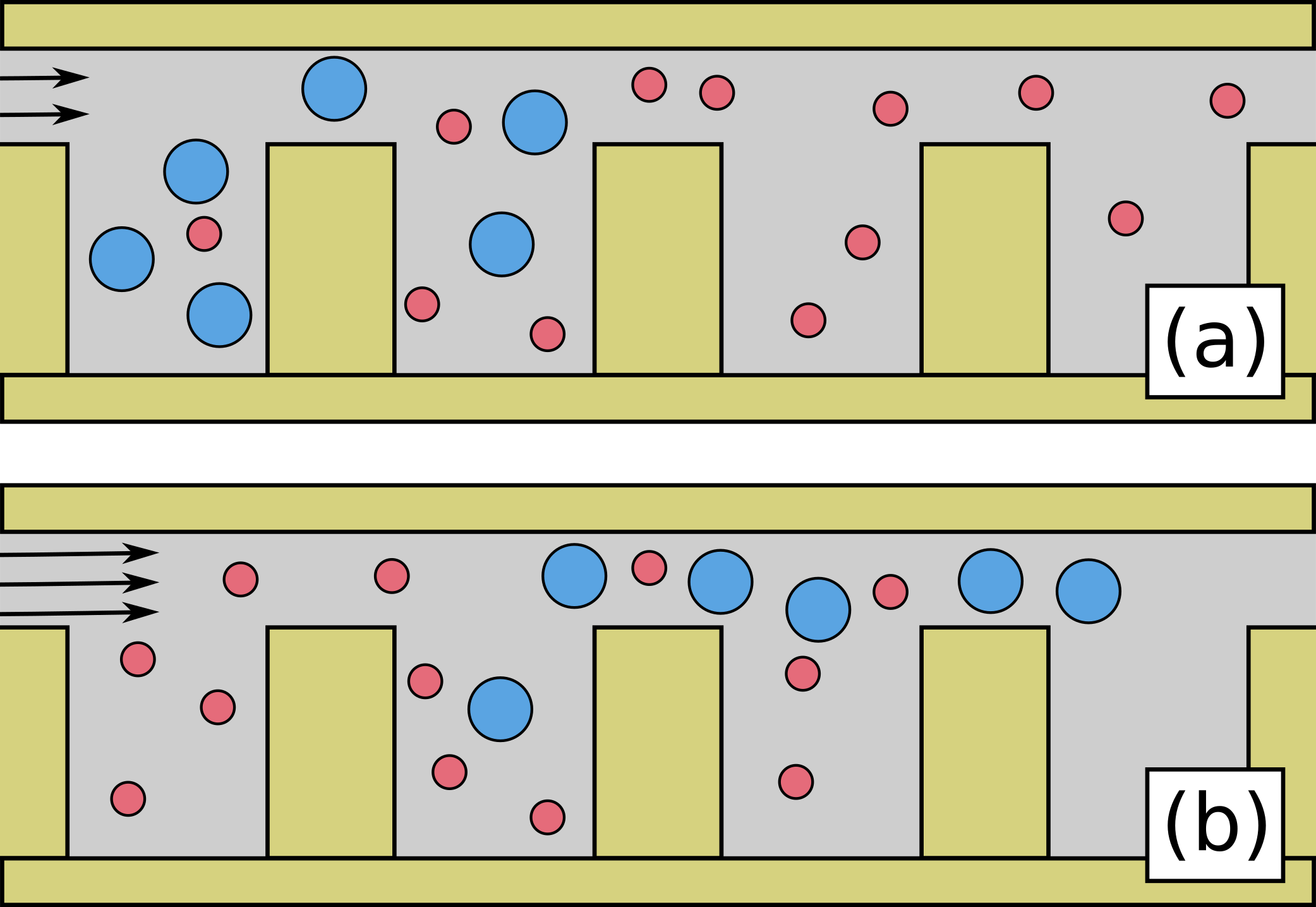}
\caption{A schematic of electrophoretic sorting of particles by size in the slit-well device. (a) A weak field causes small particles (red) to traverse the device more quickly on average. (b) A strong field causes large particles (blue) to traverse the device more quickly on average.}
\label{fig:slitwell}
\end{figure}

Micro- and nanofluidic devices (MNFDs) are tools that can be used to manipulate or detect molecules with high precision \cite{Levy2010,Dorfman2010,Abgrall2008,Gardeniers2004,Mulero2010}.
For instance, the slit-well MNFD was proposed by \citet{Han1999} as a tool for sorting otherwise free-draining polymers, such as DNA, according to chain length.
The same device has also been shown to induce separation of free-draining nanoparticles by size \cite{Cheng2008,Wang2020}.
The slit-well is operated by electrophoretically forcing analytes across a periodic array of deeper regions (wells) and shallower regions (slits) between two fixed planes (see Fig.~\ref{fig:slitwell}).
Its sorting effect has been comprehensively studied through theoretical, numerical, and experimental investigations, which have identified a variety of distinct mechanisms that are relevant in different operational regimes.
At a high level, the sorting effect depends nonlinearly on the size of the analytes and the magnitude of the applied electric field, as well as the shape and size of the device's wells and slits \cite{Cheng2008,Fu2005,Fu2006,Dorfman2010,Langecker2011,Wang2020}.
In particular, depending on the choice of these parameters, the mobility of analytes can be made either increasing or decreasing with respect to molecule size

The practical relevance of biotechnologies such as MNFDs has been stressed in the last year; for instance, \citet{Berkenbrock2020} surveyed the potential of microfluidics as a means of rapidly testing large numbers of people for COVID-19 infections.
\citet{Shepherd2021} studied a parallelized MNFD that generated scalable lipid nanoparticle formulations needed for applications in RNA therapeutics and vaccines.
Nonetheless, the design and optimization of MNFDs is often challenging, as it entails simultaneously considering the influence of many design parameters (e.g., operating voltage, solvent composition, device geometry, etc.) on multiple nonlinearly interdependent phenomena.

In many cases, important biological phenomena can fruitfully be modeled as first passage processes \cite{Chou2014}.
Moreover, in the study of MNFDs, key transport phenomena are often captured by only the first few moments of an appropriate first passage time distribution.
For example, the translocation of a polymer through a nanopore is aptly described as a first passage process, and the mean translocation time is a widely studied metric \cite{lam2019,Briggs2018,deHaan2010}.
\citet{Magill2018channels} showed that, for the special class of MNFDs with periodic geometries featuring small bottlenecks, the long-term dynamics of molecules driven through the system depend exclusively on the first and second moments of their first passage times across one subunit of the device.
The ability to focus on a handful of first passage time moments can greatly simplify the problem of characterizing and designing MNFDs.

This emphasis on the first few moments of the first passage time is of particular interest in light of a convenient mathematical property of the Smoluchowski equation\footnote{Note that the Smoluchowski equation is also variously known as the Kolmogorov forward equation, the Fokker-Planck equation, or the convection-diffusion equation, with certain names more common in certain areas of application.} that describes the motion of analytes through MNFDs.
For instance, the dynamics of nanoparticles electrophoretically driven through an MNFD can be modeled by the Smoluchowski equation as
\begin{align}
\rho_t = \nabla \cdot \left( D \nabla \rho - \mu \vec{E} \rho\right), \label{eqn:FP-time}
\end{align}
where $\rho$ is the position distribution of the particles over space and time, $D$ and $\mu$ are the diffusion and free-solution mobility coefficients of the particles, and $\vec{E}$ is the applied electric field.
In first-passage problems where the domain geometry and applied fields are time-invariant, Eqn.~\ref{eqn:FP-time} can be integrated over time to obtain the time-integrated Smoluchowski equation \cite{Redner2001}
\begin{align}
-\rho_0 = \nabla \cdot \left( D \nabla g_0 - \mu \vec{E} g_0 \right), \label{eqn:g0}
\end{align}
where $\rho_0$ is the initial condition for $\rho$.
The new field $g_0$ is defined as
\begin{align}
g_0(x,y) := \int_0^\infty \rho(x,y,t) \,dt.
\end{align}
The integral of $g_0$ in any region is the average residence time of particles in that region between initialization and absorption.
In particular, it therefore has the property that
\begin{align}
\int_\Omega g_0 \,dx = \langle \tau \rangle,
\label{eqn:tau}
\end{align}
when $\Omega$ is the entire spatial domain, $\tau$ is the stochastic first passage time of the particles to the absorbing boundary conditions, and $\langle \tau \rangle$ is the mean first passage time (MFPT).
Moreover, this formulation can be extended recursively to all higher-order moments as well.
For instance, the field $g_1$ satisfying 
\begin{align}
-g_0 = \nabla \cdot \left( D \nabla g_1 - \mu \vec{E} g_1 \right), \label{eqn:g1}
\end{align}
has the property that it integrates over the spatial domain to yield the second moment of the first passage time.
In this work, we will refer to the time-integrated Smoluchowski equation (Eqn.~\ref{eqn:g0}) as the $g_0$ equation, Eqn.~\ref{eqn:g1} as the $g_1$ equation, and the hierarchy of equations collectively as the moment equations.
A more comprehensive discussion of these moment equations can be found in standard references such as \citet{Redner2001}.

Since the first few moments of first passage time distributions are so important to MNFD phenomena, it is natural to wonder whether solving the moment equations directly might be a useful line of investigation.
In practice, however, it appears that this is rarely done.
\citet{Redner2001} shows the power of the moment equations for theoretical analysis of first passage problems, especially in the purely diffusive regime where direct analogies with electrostatics can be made.
In the context of MNFDs, \citet{Magill2018channels} showed that measuring $g_0$ approximately via particle simulations can aid in understanding the effect of design parameters on system dynamics.
Similarly, \citet{Wang2020} analyzed plots of the time-integrated particle position densities in a periodic model of the slit-well device; however, these maps were constructed in a manner subtly different from $g_0$, and in particular do not have the property of integrating to the MFPT.
The authors are unaware of other studies in which the moment equations are solved numerically towards the goal of understanding the effect of MNFD design parameters on first passage time behaviors.
Moreover, even though the Smoluchowski equation is also an important mathematical model to study first passage time problems outside biophysics \cite{Kurella2015,Iyer2015,Ma2009,Drigo2012}, we have found no examples in which the $g_0$ equation (nor any of the higher moment equations) were studied numerically in applied contexts.

A major barrier to the goal of solving the moments equations numerically in biophysics is the so-called curse of dimensionality.
That is, for most common numerical methods for partial differential equations (PDEs), the computational cost grows exponentially in the dimensionality of the underlying domain.
Thus, whereas highly effective techniques like the finite element method (FEM) can be used to solve PDEs in simple biophysical scenarios, like that of noninteracting nanoparticles, they fail when applied to the high-dimensional PDEs describing the dynamics of many-body systems such as polymers.
Indeed, particle-based simulation methods do not exhibit the curse of dimensionality, and this can be seen as a major reason for the dominance of particle simulations over PDE-based calculations in biophysics.

In this work, we attempt to resolve this barrier using a new numerical method for PDEs that does not suffer from the curse of the dimensionality.
The technique, which we refer to as the neural network method (NNM), is inspired by the success of deep learning at solving high-dimensional problems in machine learning, such as image processing and natural language processing \cite{imagenet,lecun,nlp2011}. 
A growing body of theoretical and numerical evidence suggests that it can robustly solve high-dimensional PDEs \cite{E2017,Sirignano2018,Royo2017,Avrutskiy2020,
Han2018,Hure2020,Karumuri2020,Mutuk2019,Nabian2019,
Zhang2020,Beck2019,Wei2018,Grohs2018,Jentzen2018,
Hutzenthaler2019}.
In particular, the NNM has already been used to study high-dimensional problems in biophysics \cite{Wei2018}.

The NNM has also been shown to solve parameterized problems directly across a continuous range of parameter values \cite{Sirignano2018,Hennigh2021}.
As the number of parameters increases, the problem of solving a highly parameterized PDE can exhibit yet another curse of dimensionality.
Because the neural network method shares information across parameter space, it is also able to overcome this computational challenge \cite{Berner2020,Glau2020,Khoo2021}.

Note that parameterized solutions to PDEs typically cannot be obtained using FEM, particle simulations, or similar methods.
Rather, this goal is usually accomplished using reduced order modeling (ROM) techniques \cite{Bazaz2012,Lu2021}.
ROM methods typically interpolate between a relatively small number of high-fidelity solutions computed at a handful of reference points in parameter space in order to approximate solutions at new points in parameter space.
Whereas most classical ROM methods interpolate to new parameter choices via a linear combination of the reference solutions, the NNM is intrinsically nonlinear.
Other nonlinear ROMs based on deep neural networks have been proposed in the literature \cite{Franco2021,Kasim2020,Eivazi2020,Daniel2020,Fresca2021}.
However, these methods require a database of FEM or other classical solutions be computed prior to training, whereas the NNM simultaneously solves the target PDE and acts as an ROM method over parameter space.
In addition, dealing with parameters that modify the domain geometry using classical reduced-order methods can be challenging as these are typically constructed using mesh-based approaches.
Although special ROMs can be developed for geometric parameters in some cases \cite{De2016,Heydari2001,Daniel2004}, the mesh-free nature of the NNM is intrinsically advantageous for this application \cite{Hennigh2021,Gao2021}.

\section{Problem description \label{sec:problem}}

\begin{figure*}
\includegraphics[width=\textwidth]{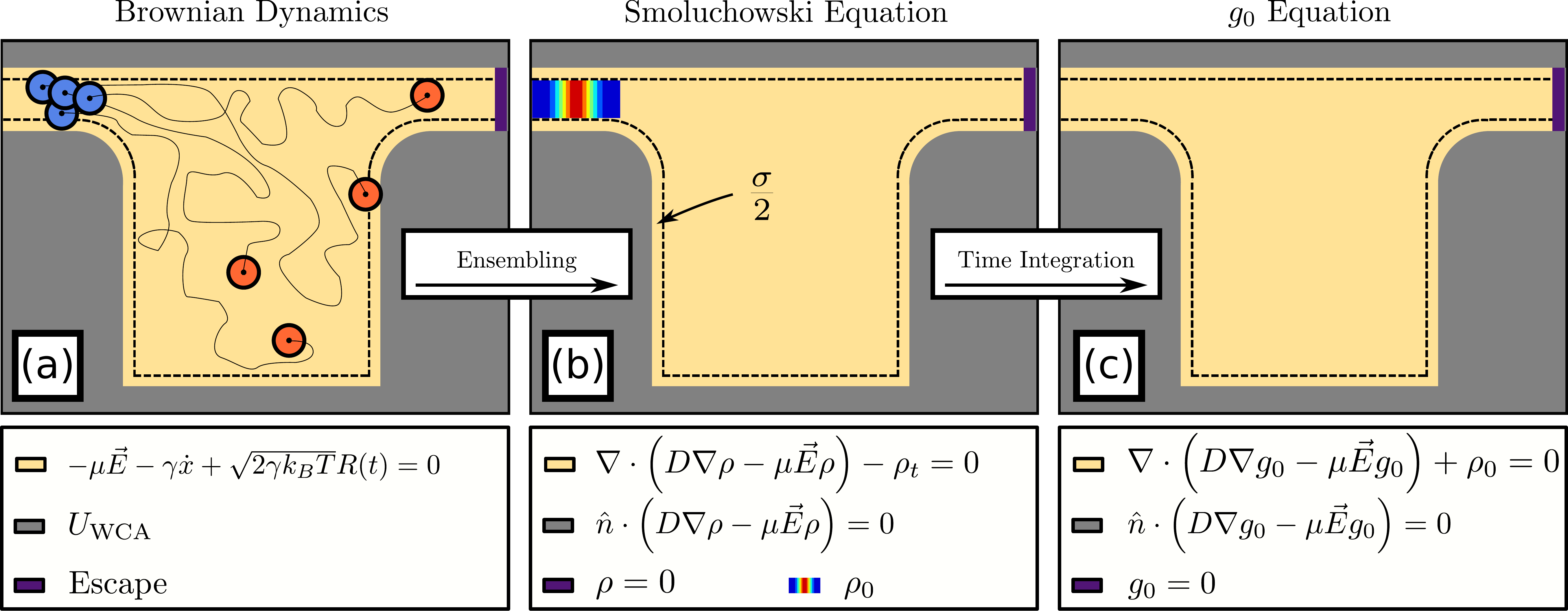}
\caption{Schematic of a single periodic subunit of the slit-well device to illustrate passage time models used in this study. 
(a) Particles are initialized in the left slit (blue particles), confined in the device via a WCA potential on grey walls, undergo Brownian dynamics in the yellow interior (trajectories denoted by black lines) until escaping device at the purple boundary.
Dotted line denotes the region of the interior that cannot be occupied by the center of mass of particles.
(b) A PDE model of the escape process where the solution $\rho$ satisfies the Smoluchowski equation in the yellow interior.
The initial Gaussian band source is located in the left slit, an absorbing boundary in the right slit (purple), and no flux conditions applied on grey walls that move inward to the dotted line to model particle size. 
(c) A PDE model of the escape process where the solution $g_0$ satisfies the $g_0$ equation (first moment) in the yellow interior region. 
An absorbing boundary is applied at the right slit wall (purple), and no flux conditions are applied on grey boundaries which move inward  the dotted line which models particle size.} 
\label{fig:problem-statement}
\end{figure*}

The primary goal of this paper is to study the effectiveness of the NNM as a tool for solving the $g_0$ equation in MNFDs by focusing on a sufficiently complicated representative device as shown in Fig.~\ref{fig:problem-statement}.
The specific problem under consideration is as follows: for an ensemble of thermal particles initially located in the left slit of one periodic subunit of the slit-well device, compute the MFPT of these particles to the right slit.
Here, the particles are driven by an electric field $\vec{E} = \lambda \vec{E}_0$ for a field strength constant $\lambda$ and a baseline electric field $\vec{E}_0$. The baseline electric field is a solution to Laplace's equation for a voltage drop of 2 across the domain.
It was obtained using the NNM in the manner described in \citet{Magill2020field}, and plots of $\vec{E}_0$ are included here in App.~\ref{app:contour}.
The particles represent nanoparticles with diameters $\sigma$, diffusion coefficients $D$, and free-solution mobilities $\mu$.
The nanoparticles are assumed to be free-draining, with $\mu$ independent of $\sigma$, so that separation by size would not occur in free solution.
Conversely, the diffusion coefficient is assumed to emerge from Stokes' law and the fluctuation-dissipation theorem, so that $D \propto \sigma^{-1}$.
For simplicity, these behaviors are implemented as $\mu=1$ and $D = \sigma^{-1}$.
The $g_0$ equation is thus reduced to 
\begin{equation}
\nabla \cdot \left( \frac{1}{\sigma} \nabla g_0 - \lambda \vec{E}_0 g_0 \right) +\rho_0  =0, \label{eqn:reduced_g0}
\end{equation}
where the particle size $\sigma$ and field strength $\lambda$ are the two free parameters, and $\vec{E}_0$ is the reference electric field.

The problem geometry is shown in Fig.~\ref{fig:problem-statement}, along with depictions of the particle-based, Smoluchowski, and $g_0$ representations of the problem.
The domain is meant to represent a single periodic subunit of the slit-well device illustrated in Fig.~\ref{fig:slitwell}.
In the particle-based model of the problem [Fig.~\ref{fig:problem-statement}(a)], an ensemble of noninteracting nanoparticles are initially located in the left slit, 
and then these particles proceed to move under a combination of thermal diffusion and electrophoretic drift until reaching the far right purple wall in the right slit.
In the Smoluchowski model [Fig.~\ref{fig:problem-statement}(b)], individual particles are eschewed, and the time evolution of the entire distribution of particle positions is modeled instead.
Here, the initial position of particles is modeled by the initial condition $\rho_0(x;\sigma)$, located in the left slit.
Finally, in the $g_0$ equation [Fig.~\ref{fig:problem-statement}(c)], the time-dependence of the Smoluchowski is accounted for implicitly by integration over all time.
Here, the initial condition $\rho_0(x;\sigma)$ now appears as a source term in the (time-independent) PDE.

In each schematic of Fig.~\ref{fig:problem-statement}, the grey regions represent physical walls.
These were modeled as short-range repulsive boundaries (in the particle model; see App.~\ref{app:appendixPS}) or no-flux boundary conditions in the continuum models [equations defined in the legends of Fig.~\ref{fig:problem-statement}(b) and (c)].
As a result of excluded volume interactions, the particle centers cannot come closer than a distance of roughly $\sigma/2$ from the repulsive boundaries.
This exclusion zone is depicted by the dashed black line in Fig.~\ref{fig:problem-statement}.
To model this in the continuum models, the no-flux boundary conditions are applied at the boundary of the exclusion zone (i.e., along the dotted black lines in Fig.~\ref{fig:problem-statement}), rather than at the nominal boundaries (i.e., along the gray walls in Fig.~\ref{fig:problem-statement}).

The nominal dimensions of the domain $\Omega_0$ are the same as those described in \citet{Magill2020field}.
In particular, the topmost and bottommost walls are a distance $2L_y = 6.25$ apart, the leftmost and rightmost boundaries are $2L_x = 10$ apart, and the curvature of the re-entrant corners is set to $R=1$ (see below).
The total horizontal lengths of the slits and the well were set equal, to $L_x$, and the slits were given a height of $h_\mathrm{slit} = L_x/4 = 1.25$.

As discussed in \citet{Magill2020field}, the standard formulation of the NNM struggles to solve problems exhibiting singularities.
For this reason, the re-entrant corners of the slit-well device geometry have been rounded (i.e., represented by circular arcs of finite curvature).
Similarly, the NNM was found to perform poorly when the initial distribution of particles was too sharp.
Instead, particles were initialized in a Gaussian band in the left slit, given by uniform distribution in $y$ multiplied by a Gaussian distribution in $x$:
\begin{align}
\rho_0 = \frac{1}{\sqrt{2 \pi} r_s h_{\mathrm{slit}}} \exp \left( \frac{-\left(x-x_s\right)^2}{2 r_s^2} \right).
\label{eqn:source}
\end{align}
Here $r_s = 0.25$ is the width of the Gaussian band in the $x$ direction, $h_{\mathrm{slit}} = L_y - y_\mathrm{slit} - \sigma$ is the height of the band in the $y$ direction, and $x_s = -L_x + 1$ is the center of the band.
Technically, $\rho_0$ requires a correction factor to be properly normalized over this bounded domain, as the Gaussian distribution in $x$ is normalized over the entire real line, but the discrepancy is numerically insignificant.

The first passage time of the particles is computed when their centers cross the rightmost boundary of the domain for the first time (purple in Fig.~\ref{fig:problem-statement}).
In the continuum models, this is represented by an absorbing boundary condition (i.e., a homogeneous Dirichlet condition).
Physically, this boundary corresponds to the interface between consecutive periodic subunits of the slit-well, and not to a physical wall.
As such, in contrast to the no-flux boundary condition on the gray walls, the placement of the absorbing boundary does not depend on $\sigma$.

As a simplifying assumption, particles were prevented from moving through the leftmost boundary of the domain.
Mathematically, this was imposed by a no-flux boundary condition.
Physically, this corresponds to the synthetic condition that particles cannot move against the direction of the imposed electric field into the previous periodic subunit of the slit-well; we will refer to this as the no-backflow condition.
The location of this no-backflow boundary condition was fixed independently of $\sigma$.

Of course, in the actual slit-well device there is always a nonzero probability of particle backflow.
The simplification was made here because it allows the $g_0$ equation to be posed in a much simpler domain (i.e., a single periodic subunit).
However, as a result of this modeling choice, there will be discrepancies between the MFPT results reported in this paper and the results of previous studies of the slit-well device (such as \citet{Cheng2008} and \citet{Wang2020}), especially at low electric field strengths.
Nevertheless, as the results in Sec.~\ref{sec:results} will indicate, the major features of the slit-well system are preserved despite the no-backflow condition.
Furthermore, the simplified model still contains several mathematical features that are expected to be common to many MNFDs and particularly difficult for the NNM to resolve: re-entrant corners, a nonuniform electric field, and nontrivial dependence on physical and geometric problem parameters.
As stated above, the purpose of this paper is to study the performance of the NNM when solving a problem with the characteristic features of a typical MNFD problem.
Certain features, such as the highly skewed geometry of the fully periodic slit-well and the singularities associated with the fully sharp re-entrant corners, are more technically challenging and relegated to future work.

\section{Methodology \label{sec:methods}}

\subsection{Neural network method \label{sec:NNM}}

The NNM implementation used for this work was similar to that previously described by \citet{Magill2020field}.
In the fixed parameter experiments (Sec.~\ref{sec:FP}), the true solution $g_0(\mathbf{x})$ of the $g_0$ equation (Eqn.~\ref{eqn:reduced_g0}) was approximated by a deep neural network $\widetilde{g_0}(\mathbf{x})$ trained to minimize a composite loss functional
\begin{align}
\mathcal{L} = \mathcal{L}_{\mathrm{PDE}} + \mathcal{L}_{\mathrm{BC}} + \mathcal{L}_{\mathrm{norm}}.
\end{align}
The first loss term consisted of 
\begin{align}
\mathcal{L}_{\mathrm{PDE}}[\widetilde{g_0}] = \int_\Omega \left( F[\widetilde{g_0}] \right)^2 dA,  \label{eqn:loss}
\end{align}
where $F$ is the operator on the righthand side of the $g_0$ equation (Eqn.~\ref{eqn:reduced_g0}).
Thus, $\mathcal{L}_{\mathrm{PDE}}[\widetilde{g_0}]$ quantifies the extent to which $\widetilde{g_0}$ satisfied the $g_0$ equation (Eqn.~\ref{eqn:reduced_g0}) throughout the domain $\Omega$. 
Note that, as discussed in Sec.~\ref{sec:problem}, $\Omega$ depends on $\sigma$.
The second loss term was defined as
\begin{align}
\mathcal{L}_{\mathrm{BC}}[\widetilde{g_0}] = \int_{\partial\Omega} \left( B[\widetilde{g_0}] \right)^2 ds,
\end{align}
where $B[\widetilde{g_0}]$ defines no-flux or absorbing boundary conditions, as appropriate, on each part of the boundary of the domain [see Fig.~\ref{fig:problem-statement}(c)].
Thus, $\mathcal{L}_{\mathrm{BC}}[\widetilde{g_0}]$ quantifies the extent to which $\widetilde{g_0}$ satisfied the BCs over the domain boundary $\partial\Omega$.

The final term was given by
\begin{align}
\mathcal{L}_{\mathrm{norm}}[\widetilde{g_0}] = \left[ \int_\Omega \left( F[\widetilde{g_0}]\right) dA \right]^2.
\end{align}
Similarly to $\mathcal{L}_{\mathrm{PDE}}$, the last loss term $\mathcal{L}_{\mathrm{norm}}$ quantifies the extent to which the approximate solution satisfies the PDE inside the domain $\Omega$.
However, whereas $\mathcal{L}_{\mathrm{PDE}}$ is a \textit{local} measure of the residual of Eqn.~\ref{eqn:reduced_g0}, $\mathcal{L}_{\mathrm{norm}}$ is a \textit{global} measure.
Specifically, $\mathcal{L}_{\mathrm{PDE}}$ is the mean of the squared residual, while $\mathcal{L}_{\mathrm{PDE}}$ is the square of the mean residual.
In theory, $\mathcal{L}_{\mathrm{norm}}$ is a redundant loss term and simply setting $\mathcal{L}_{\mathrm{PDE}}$ to zero is sufficient to ensure that $\widetilde{g_0}$ satisfies the $g_0$ equation (Eqn.~\ref{eqn:reduced_g0}).
In practice, however, training without $\mathcal{L}_{\mathrm{norm}}$ was found to produce approximate solutions that captured the shape of the true solution fairly accurately, but struggled to converge on the correct magnitude (i.e., differed from the true solution by a small multiplicative factor).

\begin{figure}
\includegraphics[width=0.98\columnwidth]{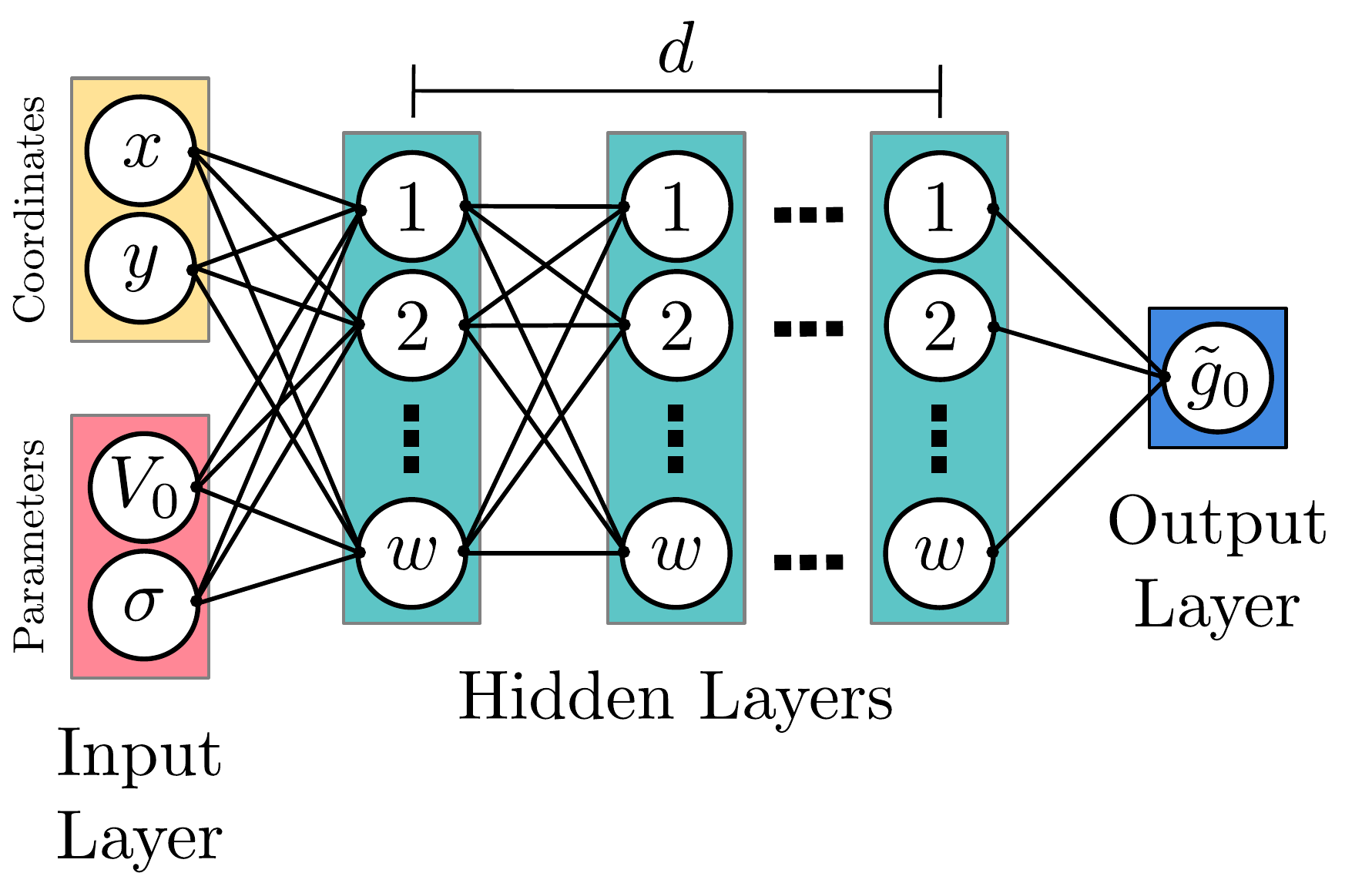}
\caption{Fully connected feedforward neural network of width $w$ and depth $d$ mapping coordinates $(x,y)$ and problem parameters $(\lambda,\sigma)$ to an output $\tilde{g}_0(x,y;\lambda,\sigma)$. At each node, a weighted sum of the incoming arrows and a bias is computed and passed through an activation function. The network's parameters are optimized such that $\tilde{g}_0(x,y;\lambda,\sigma)$ approximately satisfies the target PDE and BCs.}
\label{fig:architecture}
\end{figure}

We note that our use of $\mathcal{L}_{\mathrm{norm}}$ is similar to the normalization process used by \citet{AlAradi2018} in solving the time-dependent Smoluchowski equation.
The use of $\mathcal{L}_{\mathrm{norm}}$ can also be contrasted with previous work by \citet{Avrutskiy2020derivatives}.
There, \citet{Avrutskiy2020derivatives} showed benefits to adding redundant loss terms that encourage the solution to satisfy the derivative of the PDE operator $F'=0$ over the spatial domain.
Here, the term $\mathcal{L}_{\mathrm{norm}}$ is a redundant loss term that encourages $\widetilde{g_0}$ to agree with the \textit{integral} of the PDE operator F over the spatial domain $\Omega$.
These additional loss terms can be thought of as soft constraints on the training process, or equivalently as regularization terms constructed out of prior knowledge of the target problem. 

In Secs.~\ref{sec:vv}-4, the neural network was parameterized with respect to field strength $\lambda$, particle size $\sigma$, or both. 
Thus, the loss terms were redefined as
\begin{align}
\mathcal{L}_{\mathrm{PDE}}[\widetilde{g_0}] &=  \left\langle \int_{\Omega_\sigma} \left( F_{\lambda,\sigma}[\widetilde{g_0}] \right)^2 dA  \right\rangle_{\lambda,\sigma} ,\label{eqn:loss-pde-mean}\\
\mathcal{L}_{\mathrm{BC}}[\widetilde{g_0}] &= \left\langle \int_{\partial\Omega_\sigma} \left( B_{\lambda,\sigma}[\widetilde{g_0}] \right)^2 ds \right\rangle_{\lambda,\sigma},\label{eqn:loss-bc-mean}\\
\mathcal{L}_{\mathrm{norm}}[\widetilde{g_0}] &= \left\langle \left[ \int_{\Omega_\sigma} \left( F_{\lambda,\sigma}[\widetilde{g_0}] \right) dA \right]^2 \right\rangle_{\lambda,\sigma}.\label{eqn:loss-norm-mean}
\end{align}
The notations $\Omega_\sigma$, $F_{\lambda,\sigma}$, and $B_{\lambda,\sigma}$ indicate that the domain changes with $\sigma$, and the PDE and BC operators change with both $\lambda$ and $\sigma$.
The angled brackets indicate averages over the parameter values.
In other words, the loss used for the parameterized neural networks is identical to that used for the fixed parameter experiments, with the additional step of averaging the loss over parameter space.

Note that the electric field $\vec{E}$  was obtained by computing the electric potential $u$ using the NNM methodology of \citet{Magill2020field}.
Of course, this is not strictly necessary as $u$ could just as easily be approximated by some other method (e.g., FEM).
However, the intention was to illustrate the ease with which previously computed NNM solutions can be fed into the loss functional of new NNM solutions.
A contour plot illustrating both $u$ and $\vec{E}$ is included in App.~\ref{app:contour} (Figure~\ref{fig:contour}). 
Note that the electric potential is defined on the nominal domain $\Omega_0$ corresponding to $\sigma=0$, which differs from the actual domain $\Omega$ on which $g_0$ is defined.

All of the NNM experiments in this work were conducted with fully-connected feedforward neural networks of depth $d=3$ and width $w=50$.
The hyperbolic tangent was used for activation functions in the hidden layers, while the output layer was linear.
To solve the nonparameterized problems (Sec.~\ref{sec:FP}), the approximate solution $\widetilde{g_0}$ was constructed as
\begin{align}
\widetilde{g_0}(\mathbf{x}) &= f_{d+1} \circ f_d \circ \cdots \circ f_1 (\mathbf{x}),
\end{align}
with
\begin{align}
f_1(\mathbf{x}) &= \tanh\left( W_1 \mathbf{x} + \mathbf{b}_1 \right), \\
f_i(\mathbf{x}) &= \tanh\left( W_i f_{i-1}(\mathbf{x}) + \mathbf{b}_i \right), & i=2\ldots d, \\
f_{d+1}(\mathbf{x}) &= \textbf{W}_{d+1} f_d(\mathbf{x}) + b_{d+1},
\end{align}
where $W_1 \in \mathbb{R}^{w\times 2}$, $W_i \in \mathbb{R}^{w\times w}$ for $i = 2\ldots d$, and $W_{d+1} \in \mathbb{R}^{1\times w}$ are the network's weight matrices, while $\mathbf{b}_i \in \mathbb{R}^{w}$ for $i = 1\ldots d$, and $b_{d+1} \in \mathbb{R}$ are its biases. 

The experiments in Secs.~\ref{sec:vv}-4 considered parameterized neural networks, where one or both of the problem parameters $\lambda$ and $\sigma$ were included as additional inputs to the network.
In these cases, the networks were defined as
\begin{align}
\widetilde{g_0}(\mathbf{x};\textbf{m}) &= f_{d+1} \circ f_d \circ \cdots \circ f_1 (\mathbf{x};\mathbf{m}),
\end{align}
with $f_i$ defined as before for $i=2\ldots d+1$, but with $f_1$ adjusted to
\begin{align}
f_1(\mathbf{x}) &= \tanh\left( W_1^{(x)} \mathbf{x} + W_1^{(m)} \mathbf{m} + \mathbf{b}_1 \right),
\end{align}
where $W_1^{(x)} \in \mathbb{R}^{w\times 2}$ and $W_1^{(m)} \in \mathbb{R}^{w\times m}$, where $m$ is the length of the parameter vector $\mathbf{m}$.
In other words, the parameters ($\lambda$ or $\sigma$ or both) were concatenated to the end of the input vector of the network, and the weight matrices were adjusted accordingly.
This is illustrated schematically in Fig.~\ref{fig:architecture}.
The same approach was used by \citet{Sirignano2018} and \citet{Hennigh2021}, but can be contrasted with the recently proposed DeepONet architecture of \citet{Lu2019}.

Training was conducted in Tensorflow \cite{Tensorflow2015} version 1.15  with all unspecified hyperparameters set to their default values.
The weights were initialized using the Glorot method \cite{Glorot2010}, and biases were initialized to zero.
Weights were iteratively updated using the Adam optimizer \cite{Kingma2014} to minimize $\mathcal{L}$ with the learning rate set to $10^{-3}$ in Sec.~\ref{sec:FP}, and to $10^{-4}$ in Secs.~\ref{sec:vv}-4. 
In each iteration, the integrals in $\mathcal{L}$ were approximated by Monte Carlo sampling using the same procedure described in \citet{Magill2020field}.
Specifically, rejection sampling was applied to 10,000 nominal samples generated in the bounding box $[-5,5]\times[-3.125,3.125]$, and each smooth subunit of the boundary was randomly sampled with a linear density of about 13 points per unit length.
For the parameterized network experiments (Secs.~\ref{sec:vv}-4), the relevant problem parameters were also sampled randomly in each training iteration.
These samples were generated uniformly at random, with $\lambda$ drawn from $[5,50]$ and $\sigma$ drawn from $[0.125,0.625]$.
In particular, it was necessary to sample the parameter $\sigma$ before sampling points in $\Omega$, since the extent of $\Omega$ varies with $\sigma$.
One random parameter vector was drawn per training iteration.

The testing loss was evaluated every 1000 training iterations, using ten times more samples than during a training step. 
In the parameterized network experiments (Secs.~\ref{sec:vv}-4), the testing loss was averaged across 100 random parameter vectors.
Training was continued for a fixed number of iterations (600,000 epochs for the fixed parameter experiments, and 30,000,000 epochs for the parameterized network experiments). 
The final network was taken as that which achieved the lowest testing loss across all iterations.

\subsection{Finite element method \label{sec:FEM}}

The MFPT problem in the slit-well domain cannot be solved exactly in closed form due to the complex nature of the geometry.
Instead, approximate ground truth solutions to the problem were obtained using the finite element method (FEM).
Following \citet{Magill2020field}, the problem for $g_0$ was solved using a mixed FEM formulation implemented in FEniCS \cite{fenics}.
The electric field $\vec{E}$ included in the PDE (Eqn.~\ref{eqn:reduced_g0}) was obtained by also approximating the electric potential $u$ by a mixed FEM formulation.
As stated above, $u$ is defined on the nominal domain $\Omega_0$, whereas $g_0$ is defined on a smaller domain depending on $\sigma$.
Thus, for the FEM solutions it was necessary to first solve $u$ and $\vec{E}$ on a discretization of $\Omega_0$, project $\vec{E}$ onto a discretization of the appropriate $\Omega$, and then define the variational problem for $g_0$ on $\Omega$.

The mesh decomposition of the domain was conducted using the \verb-mshr- package in FEniCS.
The resolution parameter was set to 200, and the re-entrant corners were approximated linearly by 400 segments each. 
The same mesh parameters were used for all values of $\sigma$, and for the nominal domain, $\Omega_0$, on which $u$ was solved.

\section{Results \label{sec:results}}

\begin{figure*}
\includegraphics[width=\textwidth]{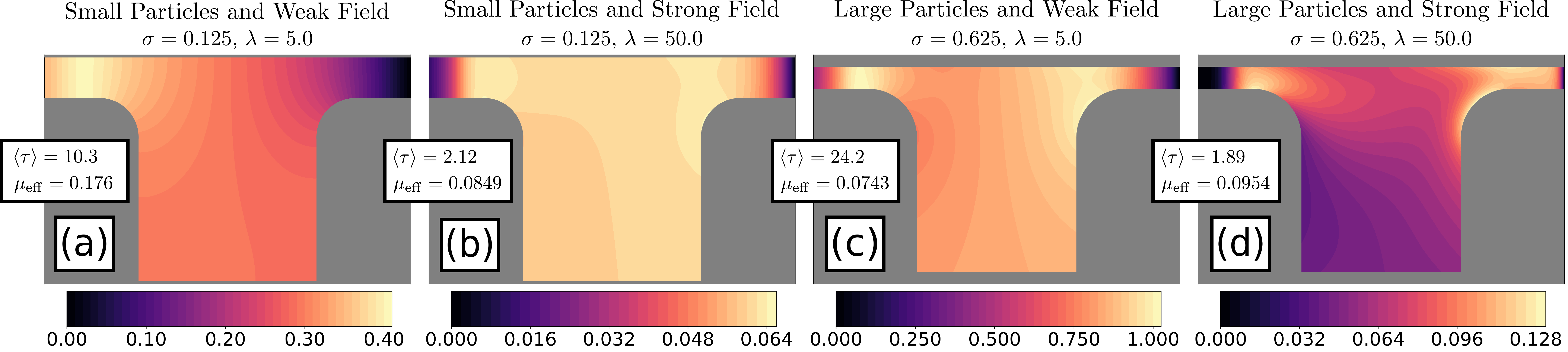}
\caption{NNM solutions to the $g_0$ equation subject to (a) a small particle size and a weak electric field, (b) a small particle size and a strong electric field, (c) a large particle size and a weak electric field, and (d) a large particle size and a strong electric field.} 
\label{fig:slit-well-soln}
\end{figure*}

This section details results obtained using the NNM to solve the $g_0$ equation modeling the MFPT of nanoparticles driven through the slit-well device (described in Sec.~\ref{sec:problem}).
The focus throughout is on the relationship between key problem parameters and observables of physical interest, where $g_0$ acts as a proxy between the two.
The first observable of interest is, naturally, the mean first passage time $\langle \tau \rangle$.
As described in Equation~\ref{eqn:tau}, $\langle \tau \rangle$ can be obtained by integrating $g_0$ over the domain $\Omega$.
Throughout this paper, the integration of $g_0$ to estimate $\langle \tau \rangle$ is accomplished using the same Monte Carlo procedure described for $\mathcal{L}_{\mathrm{PDE}}$ in Sec.~\ref{sec:methods}.

In practice, an observable of greater interest than the mean first passage time itself is the net electrophoretic mobility of the nanoparticles through the slit-well device over long timescales \cite{Cheng2008,Wang2020}.
In particular, the electrophoretic mobility is typically defined as
\begin{align}
\mu_{\mathrm{electro}} := \lim_{t\to \infty} \frac{\langle x \rangle_t}{E_c t}, \label{eqn:mu-electrophoretic}
\end{align}
where $\langle x\rangle_t$ is the ensemble average of the $x$ position at time $t$, and $E_c$ is a characteristic scale for the applied electric field strength.
It is not clear whether $\mu_{\mathrm{electro}}$ can be inferred directly from the $g_0$ problem being solved here.
Instead, the present paper will investigate a similar observable of interest, which will be called the effective mobility
\begin{equation}
\mu_{\mathrm{eff}} := \frac{ L_0/\langle \tau \rangle }{E_c} = \frac{1}{\lambda \langle\tau\rangle}. \label{eqn:mu-eff}
\end{equation}
where $L_0$ is the mean horizontal distance from $\rho_0$ to the absorbing wall. 
The characteristic field strength is chosen of the form $E_c = V_c/L_c$, where $V_c$ is a characteristic voltage drop and $L_c$ is a characteristic length scale. 
Since the overall voltage drop across the system is of order one and proportional to the field strength $\lambda$, we choose $V_c = \lambda$. 
For numerical simplicity, we also choose $L_c = L_0$, thus obtaining the final equality in Eqn.~\ref{eqn:mu-eff}.
The effective mobility is expected to exhibit similar features to the electrophoretic mobility, as both consist of characteristic particle velocities divided by characteristic electric field strengths.
A comprehensive exploration of the relationship between the two mobility definitions is left to future work.

\subsection{Characteristics of $g_0$ \label{sec:soln}}

Figure~\ref{fig:slit-well-soln} shows contour plots of $g_0$ solutions computed using the NNM, with
the corresponding estimates of $\langle \tau \rangle$ and $\mu_\mathrm{eff}$ shown in the legends.
The four subplots correspond to the four essential parameter regimes alluded to in Fig.~\ref{fig:slitwell}.
Note that the magnitude of the color scale varies across the four subplots.

First, consider the solution of $g_0$ in Fig.~\ref{fig:slit-well-soln}(a) corresponding to small particles ($\sigma=0.125$) driven by a weak field ($\lambda=5.0$).
Here, $g_0$ has a maximum in the left slit near the peak of the initial particle distribution $\rho_0$.
Naturally, since the particle positions are initialized according to $\rho_0$, the average residence time in that region is relatively high; this feature is common to all four subplots in Fig.~\ref{fig:slit-well-soln}.
Outside the left slit, $g_0$ decreases nearly monotonically from left to right, eventually reaching a value of zero on the absorbing boundary.
The shape of this function is nearly visually indistinguishable from the solution with $\sigma=0.125$ and $\lambda=0$ (not shown), and is characteristic of predominantly diffusive dynamics in all regions of the domain.

Figure~\ref{fig:slit-well-soln}(b) again shows $g_0$ for small particles ($\sigma=0.125$), but now driven by a much stronger field ($\lambda=50.0$).
In contrast with the monotonically decreasing solution in Fig.~\ref{fig:slit-well-soln}(a), in this scenario $g_0$ is relatively constant throughout most of the domain until a boundary layer near the absorber.
In fact, here $g_0$ even exhibits some minor nonmonotonic features: a shadow is evident in the bottom-left of the well, and a local maximum is attained at the entrance to the right slit.
Despite the strong driving field, the highly diffusive small particles readily explore the entirety of the well before absorption.
Drift and diffusion effects are relatively balanced in this case, with the uniformity in $x$ reflecting strongly driven motion in the horizontal direction, and the uniformity in $y$ reflecting rapid diffusion in the vertical direction.

Figure~\ref{fig:slit-well-soln}(c) shows $g_0$ for large particles ($\sigma=0.625$) driven by a weak field ($\lambda=5.0$).
Notice that the walls of the domain are shifted inward by $0.5\sigma$, reflecting the reduced area that can be occupied by the center of mass of larger particles  (Sec.~\ref{sec:problem}).
In this scenario, the smaller diffusion coefficient of the larger particles balances the weaker field, resulting in a solution that more closely resembles that in Fig.~\ref{fig:slit-well-soln}(b) than that in Fig.~\ref{fig:slit-well-soln}(a).
Here, the nonmonotonic features noted in Fig.~\ref{fig:slit-well-soln}(b) are even more pronounced.
In fact, the solution in Fig.~\ref{fig:slit-well-soln}(c) is visibly increasing from left to right across the well, in contrast to both the solutions in Figs.~\ref{fig:slit-well-soln}(a,b).
Although these nonmonotonic features are field-driven effects, the relative uniformity of $g_0$ in the vertical direction illustrates that diffusion is still an important transport mechanism in this regime.
Thus, as was the case in Fig.~\ref{fig:slit-well-soln}(b), drift and diffusion are of comparable importance; the differences between the two solutions are primarily due to the modifications to the domain geometry.

Finally, Fig.~\ref{fig:slit-well-soln}(d) shows $g_0$ computed for large particles ($\sigma=0.625$) subject to a strong electric field ($\lambda=50.0$).
Here, the shape of the solution differs significantly from those in all of Figs.~\ref{fig:slit-well-soln}(a-c).
In Fig.~\ref{fig:slit-well-soln}(d), $g_0$ takes on very small values throughout the entire well, and decreases substantially from the top of the well to its bottom.
The combination of the low diffusion coefficient and the very strong driving force causes the large particles to remain primarily streamlined in the upper region of the well as they move rapidly from $\rho_0$ to the absorber.

The MFPTs $\langle \tau \rangle$ and effective mobilities $\mu_{\mathrm{eff}}$ in the four scenarios of Fig.~\ref{fig:slit-well-soln} are consistent with the expected sorting mechanisms in each regime \cite{Cheng2008}.
When the field is strong, smaller particles have a larger $\langle \tau \rangle$ and lower $\mu_{\mathrm{eff}}$ than larger particles.
The converse is true at weak fields.
Future work should explore the relationship of $g_0$, $\langle \tau \rangle$, and $\mu_{\mathrm{eff}}$ with standard explanations for these phenomena, such as the entrance effect \cite{Cheng2008,grigoriev2002}.
The purpose of the discussion in this section was to illustrate the variety of complicated behaviours that arise in $g_0$ solutions across the different physically meaningful parameter regimes in the slit-well.
Of particular note was the significant impact of $\sigma$ on $g_0$ through geometric effects (i.e., comparing (b) to (c)).
In Sec.~\ref{sec:benchmark}, parameterized NNM solutions will be trained to interpolate nonlinearly between all four solutions in Fig.~\ref{fig:slit-well-soln}.
Ultimately, in Sec.~\ref{sec:vb} this will yield continuously differentiable mappings between both problem parameters $\lambda$ and $\sigma$ and both key physical observables $\langle \tau \rangle$ and $\mu_{\mathrm{eff}}$, thereby capturing the entirety of this rich sorting mechanism in a single numerical solution.

\subsection{Benchmarking NNM against FEM \label{sec:benchmark}}

In this section, $g_0$ will be leveraged as a proxy for the calculation of the metrics $\langle \tau \rangle$ and $\mu_{\mathrm{eff}}$. 
In practice, it is common in MNFD research and development (and scientific research more broadly) to study how such key metrics change in response to variations in the system parameters.
The simplest approach to characterizing this variation is to compute or measure the metrics independently for a large number of parameter choices. 
In Sec.~\ref{sec:FP}, the NNM is applied to precisely this task of calculating $\langle \tau \rangle$ and $\mu_{\mathrm{eff}}$ for many combinations of particle size $\sigma$ and field strength $\lambda$.

The above approach, however, requires repeated calculation of the key metrics which can be expensive when considering many independent parameters.
As discussed in Sec.~\ref{sec:intro}, the NNM can be leveraged to solve such parameterized problems directly across continuous ranges of parameter values.
The high-dimensional function $g_0(x,y;\lambda,\sigma)$ implicitly encodes $\langle \tau \rangle$ and $\mu_{\mathrm{eff}}$ as continuously differentiable functions of $\sigma$ and $\lambda$.
The NNM is used to approximate this function directly in Secs.~\ref{sec:benchmark} 2-4, for $g_0$ solutions parameterized directly by $\lambda$, $\sigma$, or both simultaneously.

\begin{figure*}
\center
\includegraphics[width=1.0\textwidth]{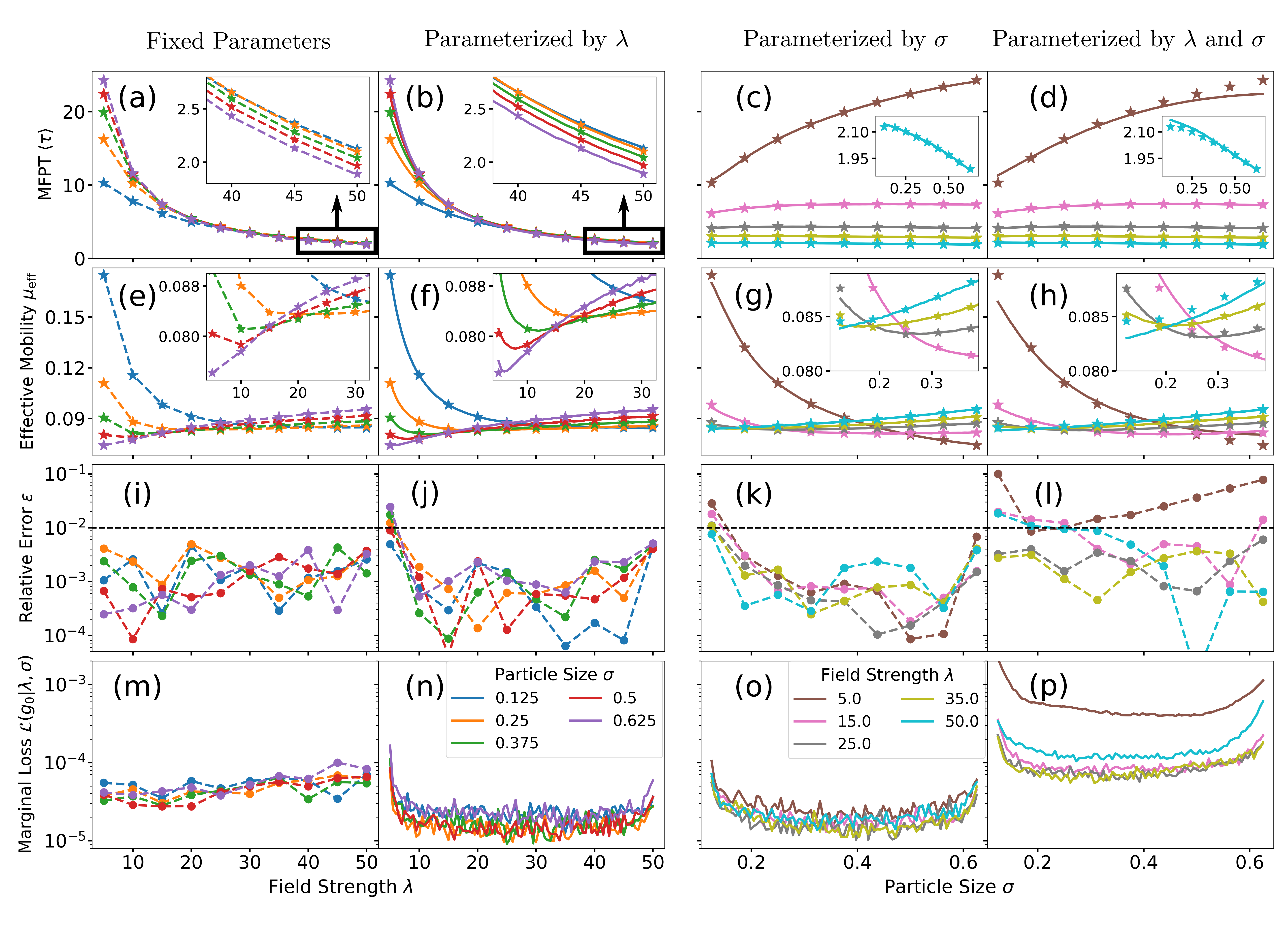}
\caption{Analysis of $g_0$ solutions computed using the NNM. 
Mean first passage times $\langle \tau \rangle$ for the NNM (a) with fixed parameters, and parameterized by (b) $\lambda$, (c) $\sigma$, and (d) both $\lambda$ and $\sigma$.
Star markers denote values obtained using FEM, and insets display behavior at high field strengths.
Effective mobilities $\mu_{\mathrm{eff}}$ for the NNM (e) with fixed parameters, and parameterized by (f) $\lambda$, (g) $\sigma$, and (h) both $\lambda$ and $\sigma$.
Star markers denote values obtained using FEM, and insets zoom in on the minimum of the curves.
Relative errors $\varepsilon$ computed against FEM for the NNM (i) with fixed parameters, and parameterized by (j) $\lambda$, (k) $\sigma$, and (l) both $\lambda$ and $\sigma$.
Dotted black line denotes 1\% error baseline computed by particle simulations.
(m) Testing loss of the NNM with fixed parameters, and marginal loss of the NNM parameterized by (n) $\lambda$, (o) $\sigma$, and (p) both $\lambda$ and $\sigma$.}
\label{fig:NNM_results}
\end{figure*}

Throughout Sec.~\ref{sec:benchmark}, four quantities are used to characterize the performance of the NNM across parameter space.
These quantities are all plotted in Fig.~\ref{fig:NNM_results}, with each column corresponding to one of the four NNM formulations discussed above.
Naturally, both the MFPT $\langle \tau \rangle$ and the effective mobility $\mu_{\mathrm{eff}}$ are included in the analysis.
These are plotted in the first two rows respectively, alongside the reference values computed using FEM.
That is, Fig.~\ref{fig:NNM_results}(a-d) shows $\langle \tau \rangle$ computed using the four NNM variations, and Fig.~\ref{fig:NNM_results}(e-h) shows the corresponding $\mu_{\mathrm{eff}}$ values.
The NNM results are indicated by lines, and the corresponding FEM results are included as stars.
Dotted lines in Fig.~\ref{fig:NNM_results}(a,e) connect values that are only computed at discrete parameter choices, whereas solid lines used everywhere else indicate values that are computed over continuous parameter ranges.
The insets in Fig.~\ref{fig:NNM_results}(a-h) provide more detail on certain subranges of the data where interesting behavior occurs.

Next, in order to quantify the accuracy of the $\langle \tau \rangle$ values obtained via the NNM, the relative error $\varepsilon$ with respect to the ground truth FEM solution is computed.
Specifically, $\varepsilon$ is defined as the relative error of $\langle \tau \rangle$ with respect to $\langle \tau \rangle_{\text{FEM}}$, i.e.,
\begin{equation}
\varepsilon = \frac{|\langle \tau \rangle - \langle \tau \rangle_{\text{FEM}}|}{\langle \tau \rangle_{\text{FEM}}},
\label{eqn:epsilon}
\end{equation}
where $\langle \tau \rangle$ and $\langle \tau \rangle_{\text{FEM}}$ are the MFPTs computed by the NNM and FEM respectively.
Note that this error metric emphasizes the use of $g_0$ as a proxy for $\langle \tau \rangle$, rather than the importance of the accuracy with which $g_0$ itself is resolved.
That said, since $\langle \tau \rangle$ is the integral of $g_0$ and integration is a linear operator, it is straightforward to see that $\varepsilon$ is bounded above by the relative $L_1$ norm of the error in $g_0$.

The relative errors $\varepsilon$ are plotted in Fig.~\ref{fig:NNM_results}(i-l).
Here, circular markers indicate the discrete parameter choices at which $\varepsilon$ was computed.
Additionally, the plots in Fig.~\ref{fig:NNM_results}(i-l) contain a dotted black line at $10^{-2}$, corresponding to a relative error of 1\%.
This is representative of a relative error threshold that is typically attainable and acceptable in MNFD research.
Indeed, App.~\ref{app:appendixPS} describes standard particle simulations that were used to approximate $\langle \tau \rangle$ independently of both the NNM and FEM.
The formulation is similar to that used in actual studies of particles in the slit-well \cite{Cheng2008,Wang2020}.
The relative error of these particle simulation results was found to be comparable to or below 1\% for all choices of parameters $\lambda$ and $\sigma$.

The final quantity included in the analysis is similar to the loss functional $\mathcal{L}$ used during the NNM training process (Eqn.~\ref{eqn:loss}).
However, the total loss provides only a single characterization of a network's performance over its entire domain.
When the NNM was used to solve parameterized $g_0$ problems, it was valuable to evaluate the relative performance of these solutions at different points in parameters space.
To this end, we defined the marginal loss $\mathcal{L}(g_0 | \lambda, \sigma)$, a parameter-dependent generalization of the total loss.
As with the true loss, $\mathcal{L}(g_0 | \lambda, \sigma)$ is the sum of the $L_2$ norms of the PDE, BC, and norm-based residuals.
However, whereas the total loss averages these quantities over all choices of $\lambda$ and/or $\sigma$ (Eqns.~\ref{eqn:loss-pde-mean}, \ref{eqn:loss-bc-mean}, and \ref{eqn:loss-norm-mean}), the corresponding terms in $\mathcal{L}(g_0 | \lambda, \sigma)$ were instead treated as functions of $\lambda$ and $\sigma$.
In the case of nonparameterized NNM solutions, the marginal loss definition simply reduces to the original total loss.

The marginal losses are plotted in Figs.~\ref{fig:NNM_results}(m-p).
The values in Fig.~\ref{fig:NNM_results}(m) correspond to NNM solutions trained at fixed parameter choices, and are thus indicated by discrete circular markers.
Conversely, since $\mathcal{L}(g_0 | \lambda, \sigma)$ can be evaluated continuously for parameterized solutions, the corresponding marginal loss values shown in Figs.~\ref{fig:NNM_results}(n-p) are indicated by solid lines sampled finely throughout the parameter space.

\FloatBarrier

\subsubsection{NNM with fixed parameters \label{sec:FP}}

This section contains a discussion of the results in Fig.~\ref{fig:NNM_results}(a,e,i,m).
Here, the NNM was applied repeatedly to solving the $g_0$ equation for fixed choices of the problem parameters: field strength $\lambda$ and particle size $\sigma$.
This NNM formulation is the same as the one used in Sec.~\ref{sec:soln}, but now applied to many more choices of the problem parameters.
Specifically, the results are shown for 10 choices of $\lambda$ uniformly spaced from $5$ to $50$ and 5 choices of $\sigma$ uniformly spaced from $0.125$ to $0.625$, with a distinct neural network used to approximate $g_0$ for each parameter combination.
The horizontal axis in Fig.~\ref{fig:NNM_results}(a,e,i,m) indicates $\lambda$, and the colors correspond to choices of $\sigma$ as per the legend in Fig.~\ref{fig:NNM_results}(n).

For small values of $\lambda$ in Fig.~\ref{fig:NNM_results}(a), $\langle \tau \rangle$ is monotonically increasing with $\sigma$, and for large values of $\lambda$ (see the inset) the opposite is true.
Moreover, the finer sampling of parameter space resolves new features that were not clear from examining only the four samples in Sec.~\ref{sec:soln}.
For instance, Fig.~\ref{fig:NNM_results}(a) shows that $\langle \tau \rangle$ decreases monotonically with $\lambda$ for each choice of $\sigma$.
In addition, the dependence of $\langle \tau \rangle$ on $\sigma$ is much stronger at low field strengths.

The same sorting behaviors can be viewed from a different perspective via $\mu_{\mathrm{eff}}$ in Fig.~\ref{fig:NNM_results}(e).
In addition, the crossover in sorting order around $\lambda \approx 25$ is better resolved by $\mu_{\mathrm{eff}}$ than  $\langle \tau \rangle$.
Indeed, in Fig.~\ref{fig:NNM_results}(e) it is clear that there is no single value of $\lambda$ for which $\langle \tau \rangle$ and $\mu_{\mathrm{eff}}$ are entirely independent of $\sigma$.
Of course, the results discussed above for $\langle \tau \rangle$ and $\mu_{\mathrm{eff}}$ are not novel, as they are consistent with published results on the slit-well device (e.g., \citet{Cheng2008}).
Rather, the purpose of this discussion is to illustrate two points.
First, that valuable information can be extracted by studying the variation of key output metrics (here, $\langle \tau \rangle$ and $\mu_{\textrm{eff}}$) as functions of the key input parameters (here, $\lambda$ and $\sigma$). 
Second, that the physical problem being studied in this paper (Sec.~\ref{sec:problem}) indeed captures essentially the same physical mechanisms expected for the actual slit-well system.

Before considering the benefits of the more ambitious parameterized NNM formulations, it is important to assess how accurately the NNM resolves $\langle \tau \rangle$ and $\mu_{\mathrm{eff}}$ when applied to the simpler task of solving $g_0$ at a single point in parameter space.
Figs.~\ref{fig:NNM_results}(a,e) suggest a good agreement between the predictions of FEM and NNM for all choices of $\lambda$ and $\sigma$.
This agreement is quantified precisely in Fig.~\ref{fig:NNM_results}(i), which shows $\varepsilon$, the relative error in $\langle \tau \rangle$.
In this plot, it appears that $\varepsilon$ is roughly independent of both $\sigma$ and $\lambda$, suggesting that the current implementation of the NNM is fairly robust throughout the problem parameter space.
This is corroborated by the testing losses $\mathcal{L}(g_0|\lambda, \sigma)$ plotted in Fig.~\ref{fig:NNM_results}(m), which are also roughly independent of the problem parameters.
Most importantly, for all choices of parameters $\lambda$ and $\sigma$ in Fig.~\ref{fig:NNM_results}(i), $\varepsilon$ is well below the 1\% error threshold indicated by the black line.
In other words, the NNM is at least as effective at resolving $\langle \tau \rangle$ as the Brownian dynamics particle simulations included in App.~\ref{app:appendixPS}.

\subsubsection{NNM parameterized by field strength \label{sec:vv}}

Whereas in Sec.~\ref{sec:FP}, 50 networks where used to obtain 50 different $g_0$ solutions, which were then integrated over their respective domains to produce 50 different $\langle \tau \rangle$ measurements, in this section only 5 networks are utilized to accomplish the same goal.
Each of these 5 networks solves $g_0(x,y;\lambda)$ for $\lambda \in [5,50]$ at a fixed choice of $\sigma$.
As in Sec.~\ref{sec:FP}, the metrics $\langle \tau \rangle$, $\mu_{\mathrm{eff}}$, $\varepsilon$, and $\mathcal{L}(g_0|\lambda,\sigma)$ are computed from the solutions; these are plotted in Fig.~\ref{fig:NNM_results}(b,f,j,n) respectively.
Comparing Fig.~\ref{fig:NNM_results}(b) to (a) and (f) to (e), it is clear that the NNM formulation parameterized by $\lambda$ recovers the same results previously obtained by solving $g_0$ independently for many different parameter choices in Sec.~\ref{sec:FP}.

One advantage of the parameterized NNM formulation is evident in the inset of Fig.~\ref{fig:NNM_results}(f).
For each choice of $\sigma$, there is a $\lambda$ value for which $\mu_{\mathrm{eff}}$ is minimal.
When computing $\mu_{\mathrm{eff}}$ only at discrete choices of the parameters [as in Fig.~\ref{fig:NNM_results}(e)], the exact location of these minima is not clear.
Instead, the results in Fig.~\ref{fig:NNM_results}(f) illustrate that the parameterized NNM formulation naturally resolves the existence of local minima, since the solution is trained continuously for all parameter values in the training domain.
The benefit of continuous mappings from problem parameters to key output metrics becomes more valuable as dimensionality of parameter space is increased (e.g., as explored in Sec.~\ref{sec:vb}).

Despite this potential advantage, the parameterized NNM formulation must be utilized with caution.
In this case, it is unclear from Fig.~\ref{fig:NNM_results}(f) alone whether the predicted minima in $\mu_{\mathrm{eff}}$ occur at the correct values of $\lambda$.
Throughout most of parameter space in Fig.~\ref{fig:NNM_results}(f), the NNM results (solid lines) appear to agree well with the reference FEM measurements (stars).
However, at $(\lambda,\sigma)=(5,0.625)$ (the leftmost point on the purple curve) the $\mu_{\mathrm{eff}}$ value predicted by the NNM deviates noticeably from the FEM measurement.
This draws into question the validity of the nearby minimum on the (purple) $\sigma=0.625$ curve.
Indeed, at this level of sampling neither the FEM nor the fixed parameter NNM [Fig.~\ref{fig:NNM_results}(e)] show any evidence that $\mu_{\mathrm{eff}}$ attains a minimum at all in this region of parameter space.
However, FEM measurements conducted on a finer sampling of $\lambda$ (not shown) confirm that there is a subtle minimum in $\mu_{\mathrm{eff}}$ near $\lambda=5.5$, for $\sigma=0.625$.
The NNM prediction shown in Fig.~\ref{fig:NNM_results}(f) places this minimum at roughly $\lambda=6$.
Regardless of the accuracy with which this feature is resolved here, this case illustrates that parameterized solutions can potentially provide clues to the existence of certain features in parameter space that would be invisible to a coarse sampling of measurements made at discrete parameter choices.

The relative error $\varepsilon$ and marginal loss $\mathcal{L}(g_0 | \lambda, \sigma)$ in Figs.~\ref{fig:NNM_results}(j,n) quantify the accuracy of the $g_0(x,y;\lambda)$ solution.
Here, both $\varepsilon$ and $\mathcal{L}(g_0 | \lambda, \sigma)$ are highest at the boundaries of the $\lambda$ training range and fairly uniform throughout the majority of the interior of the training range.
In particular, both are highest at the left boundary, $\lambda=5$.
This relationship between error and loss is similar to those studied in \citet{Magill2020field}, and provide further justification for using the (marginal) loss as an \textit{a posteriori} method for gauging the reliability of NNM solutions.

The deterioration in performance seen in Figs.~\ref{fig:NNM_results}(j,n) at the boundaries of the $\lambda$ training range can likely be attributed to the uniform Monte Carlo sampling of $\lambda$ during training.
The exact end points have very low probabilities of being sampled directly; moreover, their neighborhoods are only sampled on one side, whereas the neighborhoods of points nearer to the middle of the $\lambda$ training range are sampled thoroughly on both sides.
This could effectively lead to an under-representation of the behavior near the end points in the training loss.
Characterizing this tentative mechanism is beyond the scope of the present work.

Overall, only 3 of the 50 relative errors in Fig.~\ref{fig:NNM_results}(j) slightly exceed the 1\% error threshold.
Thus, the implementation of the parameterized NNM studied in this section meets the standard of accuracy typically attained by BD simulations.
In the regions of parameter space where the relative error was not measured directly, the marginal loss [Fig.~\ref{fig:NNM_results}(n)] provides an \textit{a posteriori} estimate of the error, suggesting that the NNM's performance is excellent except for $\lambda$ values very close to the boundaries of the training range.
Altogether, these results demonstrate that the NNM is a feasible technique for solving the $g_0$ problem over a continuous range of field strength values.
Moreover, using $g_0$ as a proxy for $\langle \tau \rangle$ and $\mu_{\mathrm{eff}}$ enables the NNM to resolve the behavior of these key output metrics continuously over the target parameter range.

 \FloatBarrier

\subsubsection{NNM parameterized by particle size \label{sec:vs}}

To expand upon the unique strengths of the NNM, this section will consider the problem of solving $g_0$ as a function of the particle size $\sigma$.
Here, since the diffusion coefficient is being modeled as $D = \sigma^{-1}$, the terms of the $g_0$ PDE depend directly on the parameter $\sigma$, just as they depend directly on $\lambda$.
However, the location of the boundaries of the slit-well domain also depend explicitly on the parameter $\sigma$ (Sec.~\ref{sec:problem}).
Thus, whereas $\lambda$ only modified the PDE terms, $\sigma$ modifies both the PDE terms and the domain geometry.
As described in Sec.~\ref{sec:intro},
it is challenging for classical reduced-order methods to deal with parameterized domain geometries.
However, this section will demonstrate that the NNM can handle geometry-modifying parameters ($\sigma$) just as easily as parameters that do not modify the domain geometry ($\lambda$).

Once again, the MFPT $\langle \tau \rangle$, effective mobility $\mu_{\mathrm{eff}}$, relative error $\varepsilon$, and marginal loss $\mathcal{L}(g_0|\lambda,\sigma)$ are computed from the NNM solutions and plotted in Fig.~\ref{fig:NNM_results}(c,g,k,o) respectively. 
Whereas the results in Fig.~\ref{fig:NNM_results}(a,e,i,m) and Fig.~\ref{fig:NNM_results}(b,f,j,n) for Secs.~\ref{sec:FP}-2 were solved and plotted as functions of $\lambda$, the results for this section are presented as functions of $\sigma$.
Specifically, each curve in Fig.~\ref{fig:NNM_results}(c,g,k,o) represents a single neural network trained over the range $\sigma \in [0.125,0.625]$ at a fixed choice of $\lambda$ (indicated by the legend in Fig.~\ref{fig:NNM_results}(o)).

The $\langle \tau \rangle$ and $\mu_{\mathrm{eff}}$ measurements in Fig.~\ref{fig:NNM_results}(c,g) indicate that the NNM parameterized by $\sigma$ recovers the same physical properties observed for the NNM with fixed parameters (Sec.~\ref{sec:FP}) and the NNM parameterized by $\lambda$ (Sec.~\ref{sec:vv}).
For the $\langle \tau \rangle$ measurements in Fig.~\ref{fig:NNM_results}(c), the $\lambda=5.0$ curve is monotonically increasing whereas the $\lambda=50.0$ curve (enhanced in the inset) is monotonically decreasing.
This implies that small particles have a lower MFPT at low field strengths, and large particles have a lower MFPT at high field strengths.
Likewise, the $\lambda=5.0$ curve for $\mu_{\mathrm{eff}}$ in Fig.~\ref{fig:NNM_results}(g) indicates that small particles are more mobile at low field strengths, whereas the $\lambda=50.0$ curve indicates that large particles are more mobile at high field strengths.
Finally, the $\lambda=25.0$ curve for $\mu_{\mathrm{eff}}$ is clearly nonmonotonic [see the inset of Fig.~\ref{fig:NNM_results}(g)].
In fact, the plots of $\mu_{\mathrm{eff}}$ make it clear that for all intermediate field strengths ($\lambda = 15,25,35$) the effective mobility is a nonmonotonic function of $\sigma$.
Altogether, the results in Fig.~\ref{fig:NNM_results}(c,g) show that the NNM recovers the same physical information whether it is trained repeatedly at independent parameters (Sec.~\ref{sec:FP}), as a continuous function of $\lambda$ (Sec.~\ref{sec:vv}), or as a continuous function of $\sigma$.

Visual comparison of the NNM results (solid lines) to the ground-truth FEM results (stars) in Figs.~\ref{fig:NNM_results}(c,g) suggests good agreement between the two through most of the parameter space.
However, the effective mobilities computed by the NNM in Fig.~\ref{fig:NNM_results}(g) deviate noticeably from the FEM results at the left endpoint $\sigma=0.125$.
Accordingly, the  relative errors and marginal losses plotted in Figs.~\ref{fig:NNM_results}(k,o) are also highest at $\sigma=0.125$.
In fact, $\varepsilon$ and $\mathcal{L}(g_0 | \lambda, \sigma)$ of the NNM solutions parameterized by $\sigma$ [Figs.~\ref{fig:NNM_results}(k,o)] exhibit the same structure previously identified (Sec.~\ref{sec:vv}) in $\varepsilon$ and $\mathcal{L}(g_0 | \lambda, \sigma)$ of the NNM solutions parameterized by $\lambda$ [Fig.~\ref{fig:NNM_results}(j,n)].
That is, $\varepsilon$ and $\mathcal{L}(g_0 | \lambda, \sigma)$ are roughly uniform for intermediate values of $\sigma$ but increase sharply near the boundaries of the training domain.
In particular, $\varepsilon$ and $\mathcal{L}(g_0 | \lambda, \sigma)$ are consistently higher at the low-$\sigma$ endpoint than at the high-$\sigma$ endpoint.

Overall, the relative error in Fig.~\ref{fig:NNM_results}(k) is well below the 1\% error threshold for most of the training range.
As was the case in Sec.~\ref{sec:vv}, at the few points where relative error exceeds 1\%, it only does so by a small amount.
The marginal loss continues to behave as an \textit{a posteriori} measure of solution accuracy, and suggests that the regions of high relative error are once again concentrated near the endpoints of the training range.
Altogether, then, these results confirm that the NNM is a feasible method for solving $g_0$ directly as a function of $\sigma$.
In particular, the parameter $\sigma$ directly changes the domain geometry in addition to modifying the terms of the PDE.
Despite this, the performance measured in this section is essentially the same as that reported in Sec.~\ref{sec:vv}, where the NNM was parameterized by the simpler parameter $\lambda$ (which does not affect domain geometry).
Thus, it appears that the NNM can handle geometry-modifying parameters just as easily as parameters that do not modify domain geometry.
This is particularly interesting given the difficulty of treating parameterized geometries with other reduced-order modeling techniques.

\subsubsection{NNM parameterized by field strength and particle size \label{sec:vb}}

\begin{figure*}[]
\center
\includegraphics[width=1.0\textwidth]{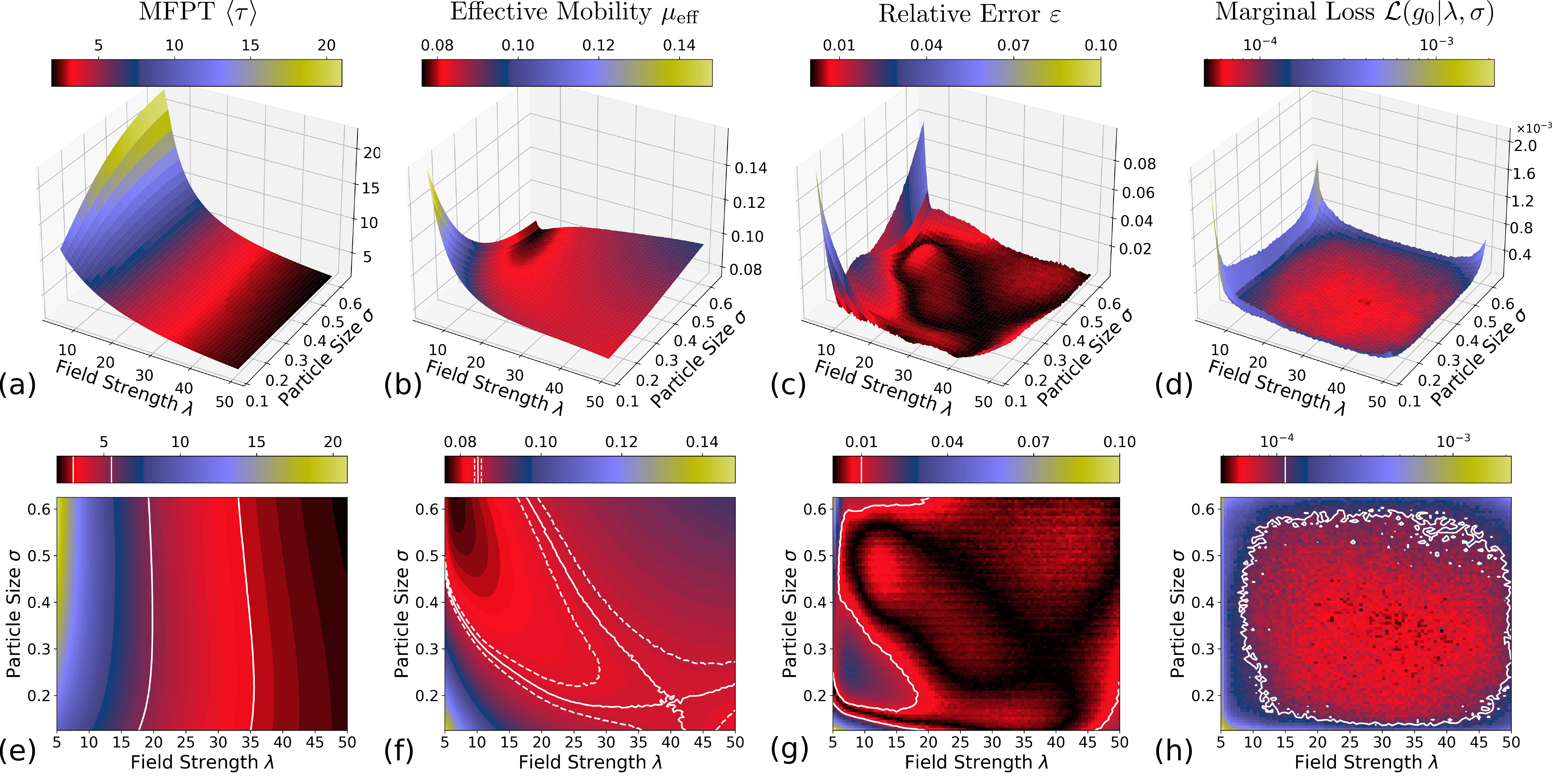}
\caption{Analysis of the $g_0(x,y;\lambda,\sigma)$ solution obtained using the NNM parameterized by both field strength $\lambda$ and particle size $\sigma$. (a,e) Mean first passage time $\langle \tau \rangle$ with white contours to show nonmonotonic sorting behavior. (b,f) Effective mobility $\mu_{\mathrm{eff}}$ with white contours to show saddle point. (c,g) Relative error $\varepsilon$ of the MFPT with white contour to denote 1\% error threshold. (d,h) Marginal loss $\mathcal{L}(g_0 | \lambda,\sigma)$ with white contour to denote testing loss $\mathcal{L}$.}
\label{fig:heatmaps}
\end{figure*}

The results shown so far have established that the NNM can robustly solve the $g_0$ equation in the slit-well MNFD (Sec.~\ref{sec:FP}), and that the method can easily be extended to produce solutions parameterized by field strength $\lambda$ (Sec.~\ref{sec:vv}) or particle size $\sigma$ (Sec.~\ref{sec:vs}).
Expanding upon this capability, in this section the NNM is used to approximate $g_0$ as a function of both $\lambda$ and $\sigma$ simultaneously (Fig.~\ref{fig:architecture}).
Specifically, a single neural network is trained to approximate the four-dimensional function $g_0(x,y;\lambda,\sigma)$ over the same parameter space previously spanned by 5 networks in Secs.~\ref{sec:vv}-3 or 50 networks in Sec.~\ref{sec:FP}.

The MFPT $\langle \tau \rangle$, effective mobility $\mu_{\mathrm{eff}}$, relative error $\varepsilon$, and marginal loss $\mathcal{L}(g_0|\lambda,\sigma)$ are computed from the NNM solution $g_0(x,y;\lambda,\sigma)$ and plotted in Figs.~\ref{fig:NNM_results}(d,h,l,p).
The lines are shown as functions of $\sigma$ and evaluated at the same choices of $\lambda$ used in Sec.~\ref{sec:vs} [indicated by the legend in Figs.~\ref{fig:NNM_results}(o)].
The $\langle \tau \rangle$ and $\mu_{\mathrm{eff}}$ values plotted in Figs.~\ref{fig:NNM_results}(d,h) closely match those in Figs.~\ref{fig:NNM_results}(c,g), demonstrating that the NNM parameterized by both $\lambda$ and $\sigma$ can resolve all the same major physical phenomena previously identified in Secs.~\ref{sec:vv}-3.

However, the accuracy of the solution $g_0(x,y;\lambda,\sigma)$ is slightly worse than that observed in the previous sections [Figs.~\ref{fig:NNM_results}(a-c,e-g)] as visible in $\langle \tau \rangle$ and $\mu_{\mathrm{eff}}$ [Figs.~\ref{fig:NNM_results}(d,h)] and quantitatively confirmed by $\varepsilon$ and $\mathcal{L}(g_0|\lambda,\sigma)$ [Figs.~\ref{fig:NNM_results}(l,p)]
This is not entirely surprising, since the four-dimensional problem here is intrinsically more difficult than the three-dimensional (Secs.~\ref{sec:vv}-3) and two-dimensional formulations (Sec.~\ref{sec:FP}) of the problem.
Moreover, the network depth and width were held constant over all experiments, and the training time was held constant for all the parameterized formulations (Sec.~\ref{sec:NNM}).
Regardless, although $g_0(x,y;\lambda,\sigma)$ appears somewhat less accurate than the solutions from previous sections, it generally still meets the target 1\% error threshold over most of its parameter training range.

An exception to this statement is presented by the results at $\lambda = 5$ [the brown lines in Figs.~\ref{fig:NNM_results}(d,h,l,p)], for which the error of $g_0(x,y;\lambda,\sigma)$ is greater than 1\% over nearly the entire $\sigma$ training range.
The marginal loss also reflects this poor performance; for $\lambda=5$, $\mathcal{L}(g_0|\lambda,\sigma)$ in Fig.~\ref{fig:NNM_results}(p) is more than an order of magnitude larger than $\mathcal{L}(g_0|\lambda,\sigma)$ from all previous experiments [i.e., those in Figs.~\ref{fig:NNM_results}(m-o)], and several times larger than the other $\mathcal{L}(g_0|\lambda,\sigma)$ curves in Fig.~\ref{fig:NNM_results}(p).
Conspicuously, the $\mathcal{L}(g_0|\lambda,\sigma)$ curves in Fig.~\ref{fig:NNM_results}(p) vary significantly with $\lambda$, whereas in Figs.~\ref{fig:NNM_results}(m-o) very little variation was observed between the different $\mathcal{L}(g_0|\lambda,\sigma)$ curves.

Of course, the results in Fig.~\ref{fig:NNM_results}(p) differ fundamentally from those in Figs.~\ref{fig:NNM_results}(m-o); whereas each curve in Figs.~\ref{fig:NNM_results}(m-o) corresponds to one or more independent networks, all the curves in Fig.~\ref{fig:NNM_results}(p) are generated by a single network.
In fact, the brown ($\lambda=5$) and blue ($\lambda=50$) curves, which exhibit the highest marginal losses in Fig.~\ref{fig:NNM_results}(p), lie directly on the boundary of the network's $(\lambda,\sigma)$ training domain.
When analysing the NNM parameterized by $\lambda$ or $\sigma$ (Secs.~\ref{sec:vv}-3), a substantial deterioration in accuracy was found to be highly localized near the boundaries of the parameter training range.
If a similar boundary effect exists here for the $g_0(x,y;\lambda,\sigma)$ solution, then the results in Figs.~\ref{fig:NNM_results}(d,h,l,p) are not representative of the solution's overall accuracy over the entire problem parameter space, as essentially half of the data shown in those plots lie on the boundary of the network's parameter training space.

To investigate this possibility, the same metrics that are shown as discrete lines in Fig.~\ref{fig:NNM_results}(d,h,l,p) are replotted in Fig.~\ref{fig:heatmaps} as continuous functions of $(\lambda,\sigma)$.
The first row [Figs.~\ref{fig:heatmaps}(a-d)] shows three-dimensional plots of the metrics over parameter space, whereas the second row shows two-dimensional contour maps [Figs.~\ref{fig:heatmaps}(e,f)] and color maps [Figs.~\ref{fig:heatmaps}(g,h)] of the same metrics.
As anticipated, the relative error $\varepsilon$ [Figs.~\ref{fig:heatmaps}(c,g)] and marginal loss $\mathcal{L}(g_0|\lambda,\sigma)$ [Figs.~\ref{fig:heatmaps}(d,h)] are only large near the boundaries of the parameter training space. 
Indeed, $\varepsilon$ in Fig.~\ref{fig:heatmaps}(g) is below the 1\% threshold (indicated by the solid white line) throughout the majority of the parameter space, confirming the suspicion that the line plots in Fig.~\ref{fig:NNM_results} provide a biased view of the $g_0(x,y;\lambda,\sigma)$ solution.

The boundary effect is particularly clear in $\mathcal{L}(g_0|\lambda,\sigma)$ in Fig.~\ref{fig:heatmaps}(d), which features a prominent convex shape.
Here, $\mathcal{L}(g_0|\lambda,\sigma)$ is consistently higher along all the edges of the training parameter space and decreases monotonically and rapidly away from the boundary.
In particular, $\mathcal{L}(g_0|\lambda,\sigma)$ is exceptionally large at the corners of the training space.

Note that the decay of marginal loss away from the boundaries of the parameter space is actually substantially sharper than it appears visually in Figs.~\ref{fig:heatmaps}(d,h).
For one, the colur scales for Figs.~\ref{fig:heatmaps}(d,h) are logarithmic.
Moreover, the color map used throughout Fig.~\ref{fig:heatmaps} is not perceptually uniform: it exhibits far more variation in color and contrast near the lower end of the scale.
These plotting choices make the subtle structure of the marginal loss more apparent, but give it the biased appearance of a gradual variation throughout the domain.
In actuality, when plotted with a linear color scale and a perceptually uniform color map, the marginal loss appears essentially flat through most of the domain.
In quantitative terms, the marginal loss varies by only a factor of 2 over the interior of the domain [i.e., within the white line in Fig.~\ref{fig:heatmaps}(h)], whereas it grows by roughly an order of magnitude near the edge of the domain (i.e., outside the white line).

As expected, the relative error $\varepsilon$ [Figs.~\ref{fig:heatmaps}(c,g)] is closely tied to the marginal loss $\mathcal{L}(g_0|\lambda,\sigma)$.
Relative error is uniformly low in the interior of the parameter space (roughly $(\lambda,\sigma) \in [15,45] \times [0.2,0.6]$), corresponding to the flat interior of $\mathcal{L}(g_0|\lambda,\sigma)$.
Additionally, near the two corners at $\lambda=5$ where $\mathcal{L}(g_0|\lambda,\sigma)$ is largest, $\varepsilon$ also attains its highest values, approaching 10\%.
There is also a small peak in $\varepsilon$ at the $(\lambda,\sigma)=(50,0.125)$ corner, corresponding to an equally small peak in $\mathcal{L}(g_0|\lambda,\sigma)$ at the same corner.
Surprisingly, although $\mathcal{L}(g_0|\lambda,\sigma)$ exhibits a clear peak at the $(\lambda,\sigma)=(50,0.625)$ corner, $\varepsilon$ does not.
Therefore, the marginal loss $\mathcal{L}(g_0|\lambda,\sigma)$ once again appears to act as a conservative \textit{a posteriori} estimator of relative error $\varepsilon$: high relative error occurs near regions of high marginal loss, although high marginal loss does not always imply high relative error.

As noted above, the performance of the $g_0(x,y;\lambda,\sigma)$ solution deteriorates even more significantly at the corners of the parameter training space than on its edges.
This is more complicated than the boundary effect discussed for the solutions parameterized by just $\lambda$ or $\sigma$, and can be accounted for by extending the postulated mechanism from Secs.~\ref{sec:vv}-3.
There, it was argued that the deterioration in performance arises because the stochastic sampling used during training under-represents boundary points: whereas the neighborhoods of interior points are thoroughly sampled on all sides, this is not true for boundary points.
In the one-dimensional parameter training spaces considered in Secs.~\ref{sec:vv}-3, the boundary of the parameter training range consists only of the two end points, which are equally affected by the sampling bias.
In the two-dimensional parameter training space considered here, however, the corners and the edges of the boundary are under-represented to different extents by the stochastic sampling process.
Whereas parameter values on the edges of the domain only have 50\% as many neighboring points inside the training space as interior points, parameter values on the corners have only 25\% as many.
This tentatively explains why performance is so much worse at corners of the parameter training space than it is on the edges.

If this mechanism extends to higher dimensional parameter spaces, it may eventually prove to be a dominant source of error: for instance, the corners of an $n$-dimensional hypercube have only $1/2^n$ as many neighbors inside the training space as interior points, and the number of boundary segments (corners, edges, faces, $\ldots$) grows rapidly with $n$.
In fact, the fraction of a parameter space lying within a given distance of its boundary also increases with dimensionality.
Altogether, these observations suggest the need for further investigation into this boundary effect, its possible connection to Monte Carlo sampling of the loss during training, and methods (such as low-discrepancy sampling methods \cite{Chen2019,Mishra2021}) for resolving the problem.

In contrast to the line plots in Fig.~\ref{fig:NNM_results}, the plots in Fig.~\ref{fig:heatmaps} highlight the richness of information available through the $g_0(x,y;\lambda,\sigma)$ solution compared to the solutions parameterized only by $\lambda$ (Sec.~\ref{sec:vv}), $\sigma$ (Sec.~\ref{sec:vs}), or neither (Sec.~\ref{sec:FP}).
For instance, although the NNM solutions parameterized by $\lambda$ or $\sigma$ (Secs.~\ref{sec:vv}-3) suggested a nonmonotonic dependence of $\langle \tau \rangle$ and $\mu_{\mathrm{eff}}$ with respect to $\sigma$ for certain values of $\lambda$, they did not provide sufficient information to estimate the exact range of $\lambda$ over which this behavior persists.
Just as the NNM solutions parameterized by $\lambda$ or $\sigma$ were shown in Secs.~\ref{sec:vv}-3 to be more helpful than the fixed parameter solutions (Sec.~\ref{sec:FP}) in localizing one-dimensional critical points, so is the NNM solution parameterized by both $\lambda$ and $\sigma$ more useful for delineating the nonmonotonic regions of parameter space.

The range of nonmonotonic behavior can be estimated visually from $\langle \tau \rangle$ in Fig.~\ref{fig:heatmaps}(e).
In that plot, whenever a given contour line crosses the same value of $\lambda$ twice, $\langle \tau \rangle$ depends nonmonotonically on $\sigma$.
However, because the contours of $\langle \tau \rangle$ are nearly parallel to the contours of $\lambda$, it is difficult to visually identify the regions of nonmonotonicity using this vertical line test.
In contrast, the $\mu_{\mathrm{eff}}$ plots in Figs.~\ref{fig:heatmaps}(b,f) are much easier to interpret.
Again using a vertical line test, it is easy to see that nonmonotonic dependence of $\mu_{\mathrm{eff}}$ on $\sigma$ is present at voltages as low as $\lambda \approx 10$.
In fact, $\mu_{\mathrm{eff}}$ is doubly nonmonotonic with respect to both $\lambda$ and $\sigma$ in the large-$\sigma$, low-$\lambda$ range [black region in Figs.~\ref{fig:heatmaps}(b,f)].
Although the same trends were suspected from the solutions discussed in Secs.~\ref{sec:vv}-3, $g_0(x,y;\lambda,\sigma)$ resolves the features more completely.

Despite the usefulness of $g_0(x,y;\lambda,\sigma)$ for resolving critical points, the solution predicts a false saddle point in $\mu_{\mathrm{eff}}$ at $(\lambda,\sigma)\approx(40, 0.2)$.
The contour passing through this saddle point is highlighted by the white solid line in Fig.~\ref{fig:heatmaps}(f).
Additional FEM results (not shown) confirm that there is no saddle point anywhere in the parameter space under consideration. 
This error can be attributed to the fact that the true $\mu_{\mathrm{eff}}$ changes extremely little in the high-$\lambda$, low-$\sigma$ region of the domain.
For illustration, the two dotted white lines in Fig.~\ref{fig:heatmaps}(f) indicate contours for $\mu_{\mathrm{eff}}$ values 1\% greater and smaller, respectively, than the value of $\mu_{\mathrm{eff}}$ on the solid white line passing through the saddle point.
As evident on the color bar above the plot in Fig.~\ref{fig:heatmaps}(f), this $\pm 1\%$ range corresponds to a very small fraction of the total variation of $\mu_{\mathrm{eff}}$ over the domain.
Despite this, the area between the dotted white lines account for roughly 25\% of the total parameter training space, demonstrating that $\mu_{\mathrm{eff}}$ is extremely flat throughout this entire region.
Indeed, this is how it is possible for the relative error [Fig.~\ref{fig:heatmaps}(g)] in the proximity of the false saddle point to remain at 1\% or below.
In other words, although the presence of a false saddle point is a noticeable qualitative error, it corresponds to a very small quantitative error in the key output metrics $\langle \tau \rangle$ and $\mu_{\mathrm{eff}}$.
In fact, the visual appearance of the saddle point in Fig.~\ref{fig:heatmaps}(f) is intentionally accentuated by the choice of color map, as discussed for the marginal loss above.
It is quite feasible that an error of such small magnitude could be resolved simply by increasing network capacity and/or training time.

Still, the question arises of whether and how the NNM can be used reliably in applications where these types of incorrect or ill-conditioned features may occur.
The marginal loss $\mathcal{L}(g_0|\lambda,\sigma)$ provides one possible resolution to this concern.
The region of increased $\mathcal{L}(g_0|\lambda,\sigma)$ near $(\lambda,\sigma)=(50,0.125)$ in Fig.~\ref{fig:heatmaps}(h) coincides fairly closely with the right half of the saddle point in Fig.~\ref{fig:heatmaps}(f).
Thus, $\mathcal{L}(g_0|\lambda,\sigma)$ correctly reflects that the solution is less reliable in this region.
If the baseline FEM solutions had not been available, the behavior of $\mathcal{L}(g_0|\lambda,\sigma)$ could have been used to draw into question the validity of the predicted saddle point.
In other words, the marginal loss $\mathcal{L}(g_0|\lambda,\sigma)$ once again acts as a conservative \textit{a posteriori} error estimator, bolstering the robustness of the NNM.
Future work should elaborate on what quantitative predictions of solution quality can be based on the marginal loss, along the lines of the investigations in \citet{Magill2020field}.
In the interim, we propose using the mean marginal loss (which is in fact the total loss, given by Eqn.~\ref{eqn:loss}) as an approximate threshold between regions of relatively high and low expected accuracy.
This is indicated in Fig.~\ref{fig:heatmaps}(h) by the solid white line.
As expected, the region of low expected accuracy is concentrated near the boundary, and in particular includes part of the false saddle point.

In fact, the marginal loss $\mathcal{L}(g_0|\lambda,\sigma)$ as defined here is likely a suboptimal tool for the detection of false critical points in parameter space, because it does not directly measure gradient information with respect to $(\lambda,\sigma)$.
Rather, it is only indirectly sensitive to the error in the shape of $\langle \tau \rangle$ and $\mu_{\mathrm{eff}}$ insofar as it emerges from errors in the shape of $g_0(x,y;\lambda,\sigma)$.
For applications in which the localization of ill-conditioned critical points is of interest, modified loss functions that incorporate the derivatives of the target PDE with respect to $\lambda$ and $\sigma$ (e.g., like those explored by \citet{Avrutskiy2020derivatives}) might be more relevant error estimators.
This notion illustrates the potential benefits of customizing the NNM for specific PDEs and research questions, just as flux- or energy-conserving numerical methods are preferred for applications where those features are particularly important.

In summary, the results in this section demonstrate that the NNM can produce a robust approximation to the $g_0(x,y;\lambda,\sigma)$ solution.
Here, $g_0(x,y;\lambda,\sigma)$ enables higher dimensional visualization of $\langle \tau \rangle$ and $\mu_{\mathrm{eff}}$ over $\lambda$ and $\sigma$, resolving features in parameter space more accurately and completely than the solutions parameterized by only $\lambda$ or $\sigma$.
Furthermore, $g_0(x,y;\lambda,\sigma)$ accurately predicts the magnitude of $\langle \tau \rangle$ and $\mu_{\mathrm{eff}}$ to within the 1\% error threshold simultaneously over the majority of the parameter training space.
Although $g_0(x,y;\lambda,\sigma)$ exhibits some regions of high relative error, $\mathcal{L}(g_0|\lambda,\sigma)$ once again provides a robust \textit{a posteriori} estimator of the solution's reliability throughout parameter space.

\section{Conclusions}

This work investigated the use of the neural network method to solve a parameterized time-integrated Smoluchowski equation describing nanoparticle passage through the slit-well microfluidic device.
The $g_0$ solutions were solved for a variety of fixed choices of field strength $\lambda$ and particle size $\sigma$ using both the NNM and a standard FEM implementation.
Additionally, the NNM was used to solve the equation directly as a function of $\lambda$ and/or $\sigma$.
Mean first passage time $\langle \tau \rangle$ and effective mobility $\mu_{\mathrm{eff}}$ were studied as the primary output metrics of interest, with relative error $\varepsilon$ and marginal loss $\mathcal{L}(g_0 | \lambda,\sigma)$ used to characterize solution performance.

The qualitative examinations of $g_0$ in Sec.~\ref{sec:soln} revealed a wide variety of functional behaviors over the region of $(\lambda,\sigma)$ parameter space studied here.
The four primary regimes underlying nanoparticle sorting in the slit-well (i.e., low and high fields for small and large particles) correspond to four significantly different $g_0$ solution types, each reflecting different interplays of drift, diffusion, and geometry.
This highlights the challenging nature of the parameterized PDE problem studied in this work.
Additionally, this analysis suggested that the $g_0$ solutions themselves may encode interesting and useful qualitative information about biophysical processes.
Future work should examine how information encoded in $g_0$ may be complementary to qualitative information derived from stochastic particle trajectories.

Of course, qualitative insights aside, the most salient feature of $g_0$ is that it integrates to yield the mean first passage time, $\langle \tau \rangle$.
As noted, although $\langle \tau \rangle$ is a quantity of widespread interest in all first passage problems and is relevant to many MNFD design problems, it appears that numerical solutions of $g_0$ have rarely been leveraged for such applications.
The results of this paper support that $g_0$ may be an undervalued tool in computational biophysics.

Although $g_0$ can be computed using many methods, such as FEM or particle simulations, this work focused on resolving $g_0$ using the NNM.
When applied to fixed choices of problem parameters, the NNM consistently estimated $\langle \tau \rangle$ with errors below 1\%.
In particular, the NNM values were at least as accurate as typical particle simulations, which are the most common tool for studying first passage problems in biophysics.
However, runtime was not compared in this work, and should be a major focus of future investigations.

The main appeal of the NNM is the unique ease with which it can be applied to parameterized $g_0$ problems.
Via integration of $g_0$, these solutions yield a direct mapping from key problem inputs (e.g., $\lambda,\sigma$) to key problem outputs (e.g., $\langle \tau \rangle, \mu_{\mathrm{eff}}$).
This is particularly appealing for the application of MNFD research, where essential phenomena often depend nontrivially on the coupling of many system parameters.
The results in the current work demonstrate that the NNM can learn accurate approximations of $g_0$ parameterized by $\lambda$, $\sigma$, or both, all using a modest network size and even without careful hyperparameter optimization.
Whereas classical ROM techniques typically require special considerations to handle geometry-modifying parameters like $\sigma$, the NNM was found to resolve $g_0(x,y;\sigma)$ just as easily as $g_0(x,y;\lambda)$.
As discussed, parameterized solutions can be quite useful in characterizing entire regions of parameter space.

Although the NNM is expected to perform well on highly parameterized PDEs, the careful error analysis presented in the current study revealed several points of caution for future efforts in this direction.
Firstly, all parameterized solutions studied here exhibited a deterioration in accuracy near the boundaries of their parameter training space.
Nonetheless, the predicted values of $\langle \tau \rangle$ were still mostly within the 1\% margin of error.
Moreover, the marginal loss functional $\mathcal{L}(g_0 | \lambda,\sigma)$ proposed here was found to act as a conservative \textit{a posteriori} estimator of the solution accuracy throughout parameter space.
That is, in any region where the relative error $\varepsilon$ of the predicted $\langle \tau \rangle$ was large, $\mathcal{L}(g_0 | \lambda,\sigma)$ was also large.

The second point of caution that must be considered when applying the NNM to parameterized PDEs concerns the interpretation of key features, such as critical points, that are identified using these solutions.
For instance, in Sec.~\ref{sec:vb}, the NNM solution exhibited an erroneous saddle point in a flat region of $\mu_{\mathrm{eff}}$, which was an artifact that arose due to the ill-conditioning of the gradients of $\mu_{\mathrm{eff}}(\lambda,\sigma)$.
In fact, plots of $\varepsilon$ showed no indication of errors in this region, as the mistake only manifested in the curvature of the mapping.
However, once again the marginal loss $\mathcal{L}(g_0 | \lambda,\sigma)$ did indicate that the NNM solution lost fidelity in this region of parameter.

In summary, the parameterized NNM solutions were generally accurate far from the training boundaries, and $\mathcal{L}(g_0 | \lambda,\sigma)$ provided robust regions of confidence.
Altogether, these results highlight the specific appeal of the NNM as a method for studying parameterized first passage problems via the time-integrated Smoluchowski model.
We hope this work prompts further investigation into the use of $g_0$ with or without the NNM, and into the relationship of  $\langle \tau \rangle$ and $\mu_{\mathrm{eff}}$ to more standard MNFD metrics.
Regarding the application of the NNM to such problems, future work should address technical challenges such as singularities posed by sharp corners, training difficulties for highly skewed geometries, and achieving competitive runtime.

\section*{Acknowledgments}

The authors gratefully acknowledge funding from Mitacs under the Accelerate Entrepreneur program (Ref. IT21168) and from Smart Computing for Innovation (SOSCIP Ref. 3-076).
H.W.d.H. also acknowledges funding from the Natural Sciences and Engineering Research Council (NSERC) in the form of Discovery Grant 2020-07145.
M.M. also acknowledges funding from the Ontario Graduate Scholarship (OGS) Program and the Vector Institute Postgraduate Affiliate Program.

\appendix
\section{COMPUTING MEAN FIRST PASSAGE TIMES WITH PARTICLE SIMULATIONS \label{app:appendixPS}}

This section contains a description of standard Brownian dynamics (BD) simulations used to measure the mean first passage times of nanoparticles traversing the slit-well microfluidic device (Sec.~\ref{sec:problem}).
The BD simulations were initialized with $N = 100,000$ particles placed according to the distribution $\rho_0$ (Eqn.~\ref{eqn:source}).
The position of the $i$th particle $\vec{x}_i$ was updated according to the discretized BD equation
\begin{align}
\frac{\Delta \vec{x}_i}{\Delta t} &= \sqrt{\frac{2 D}{\Delta t}} \vec{R}(t) + \frac{q \lambda}{\gamma} \vec{E}_0 + \frac{1}{\gamma} \vec{F}_{\mathrm{WCA}}. \label{eqn:bd}
\end{align}
In Eqn.~\ref{eqn:bd}, the particle properties are the diffusion coefficient $D$, the friction coefficient $\gamma$, and the particle charge $q$.
As noted in Sec.~\ref{sec:problem}, both $q$ and $\gamma$ were set equal to the particle diameter $\sigma$, to capture free-draining behavior.
The diffusion coefficient $D$ was set to $1/\sigma$ and the time step was set $\Delta t=10^{-5}$
The term $\vec{R}(t)$ in Eqn.~\ref{eqn:bd} is a random driving force representing the thermal motion of an implicit solvent which was sampled from a uniform distribution of mean 0 and variance 1.

Rather than representing the interactions between particles and walls as perfectly rigid, the walls were implemented using a short-range repulsive shifted WCA force
\begin{align}
\vec{F}_{\mathrm{WCA}} = - \nabla U_{\mathrm{WCA}},
\end{align}
with
\begin{equation}
U_\text{WCA}\left( r_i \right) = 
  \begin{cases}
    4\epsilon \left[ \left(\frac{\sigma_0}{r_i}\right)^{12} - \left(\frac{\sigma_0}{r_i}\right)^6 \right] 
    + \epsilon, & r_i < r_\text{cut} \\
    0, & r_i \geq r_\text{cut}.
  \end{cases}
\end{equation}
where $r_i$ is the minimum distance from particle $i$ to the nearest reflective wall minus a distance $r_{\mathrm{shift}} = 0.5 (\sigma - \sigma_0)$.
Here $r_{\mathrm{shift}}$ corresponds to the radius of the hard core of the particle, whereas $\sigma_0=0.125$ is the length over which the surface of the particle is partially compressible.
The potential is zero beyond a cutoff distance $r_\mathrm{cut} = 2^{1/6} \sigma_0$, so that if the center of the particle is farther than a distance $r_{\mathrm{shift}}+r_\mathrm{cut}$ from the wall there is no interaction.
The energy scale of the repulsive interaction was set to $\epsilon=0.125=\sigma_0$.

Although this type of model is commonly used for particle-wall interactions, due to its improved numerical stability relative to perfectly rigid interactions, it introduces a small difference between the underlying physics of the BD simulations and the PDE models being solved in this work.
For this reason, the MFPTs determined using particle simulations should not be expected to agree exactly with those obtained using the NNM and FEM methods, even in the limit of small $\Delta t$ and large $N$.
Nonetheless, as our results corroborate, the effect of this difference between the models is small.

\begin{figure}
\includegraphics[width=0.85\columnwidth]{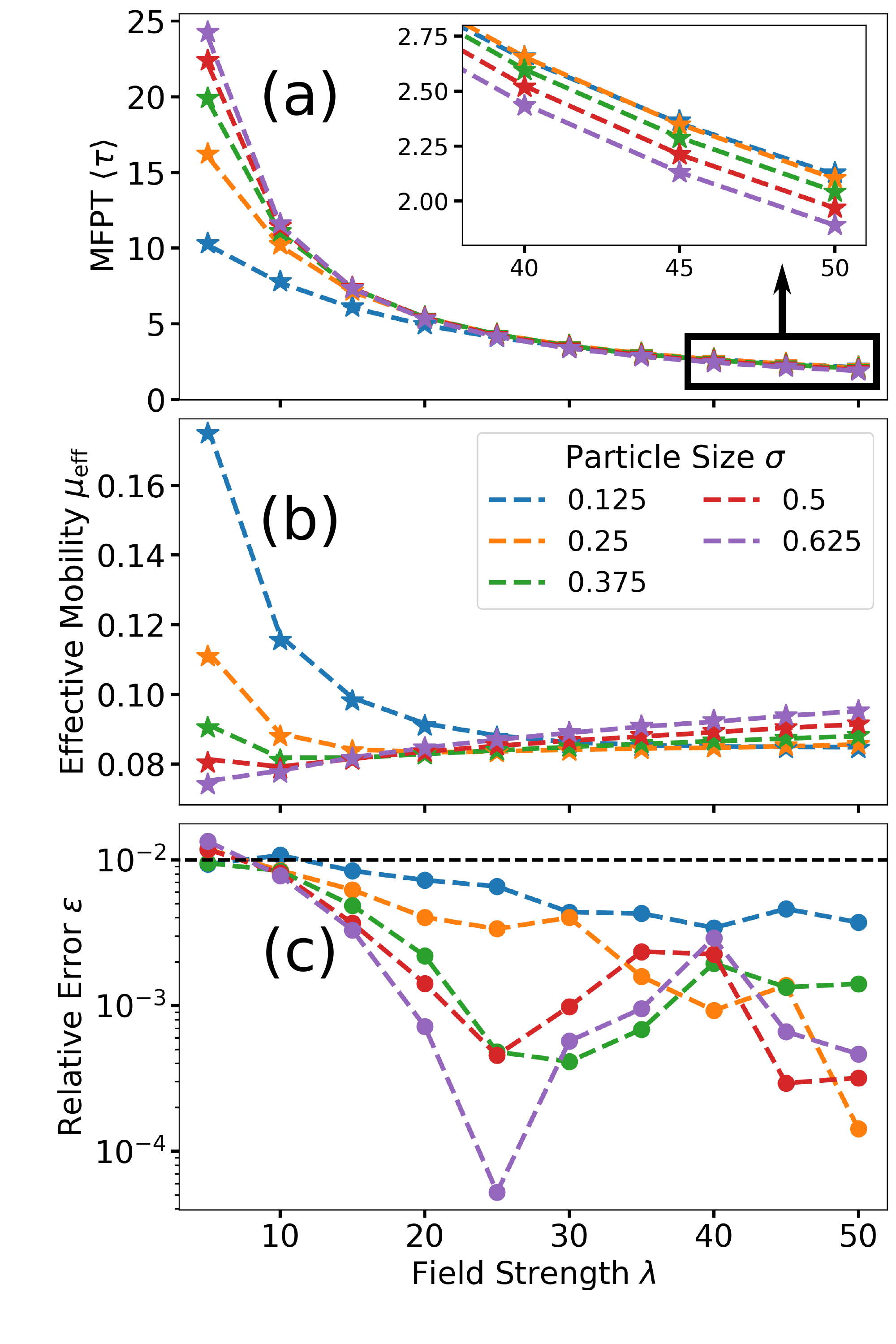}
\caption{Passage time properties of particles escaping the slit-well model computed using Brownian dynamics simulations. Star markers denote values obtained via FEM. (a) Mean first passage time $\langle \tau \rangle$. (b) Effective mobility $\mu_{\mathrm{eff}}$. (c) Relative error of $\langle \tau \rangle$ computed against the ground truth FEM solution.}
\label{fig:bd}
\end{figure}

The term $\vec{E}_0$ in Eqn.~\ref{eqn:bd} corresponds to the baseline electric field in the slit-well domain (denoted by red in Fig.~\ref{fig:contour}).
This was solved for a voltage drop of 2 units from the leftmost to rightmost boundaries, as in \citet{Magill2020field}.
The net electric field strength was set by the parameter $\lambda$.
$\vec{E}_0$ used here was the same one described in Sec.~\ref{sec:FEM}.
As shown in \citet{Magill2020field}, particle simulations conducted using an electric field solved with the NNM are nearly statistically indistinguishable from those conducted using a field solved with the FEM, so long as the NNM electric field exhibits a sufficiently small loss.
The purpose of the present study is not to replicate this result, but to explore the computational advantages of the NNM over other techniques in parameterized problems.
Thus, the particle simulations are conducted using the FEM electric field, which is taken as the reference ground truth.

Parallel to the analysis conducted in Sec.~\ref{sec:benchmark}, the mean first passage time $\langle \tau \rangle$ and  effective mobility $\mu_{\mathrm{eff}}$ are computed using the BD simulations for various choices of field strength $\lambda$ and particle size $\sigma$. 
These values are plotted with dashed lines in Fig.~\ref{fig:bd}(a) and (b) with star markers to denote $\langle \tau \rangle$ and $\mu_{\mathrm{eff}}$ values obtained by FEM.
Note, $\langle \tau \rangle$ and $\mu_{\mathrm{eff}}$ are only solved for the same discrete choices of $\lambda$ and $\sigma$ that are also computed using FEM.

In addition, the relative error $\varepsilon$ is computed using Eqn.~\ref{eqn:epsilon} where $\langle \tau \rangle$ and $\langle \tau \rangle_{\text{FEM}}$ are the MFPTs computed by BD and FEM respectively.
The values are plotted in Fig.~\ref{fig:bd}(c) with circular markers denoting the parameter choices where the relative error was computed.
All of the relative errors in Fig.~\ref{fig:bd}(c) fall below 2\%, with majority of the values being within 1\% error.
This establishes a 1\% error baseline against the groundtruth MFPT values computed by FEM, for which to benchmark the performance of the NNM.

\section{CONTOUR PLOTS OF ELECTRIC POTENTIAL AND FIELD \label{app:contour}}

\begin{figure}[h]
\includegraphics[width=0.85\columnwidth]{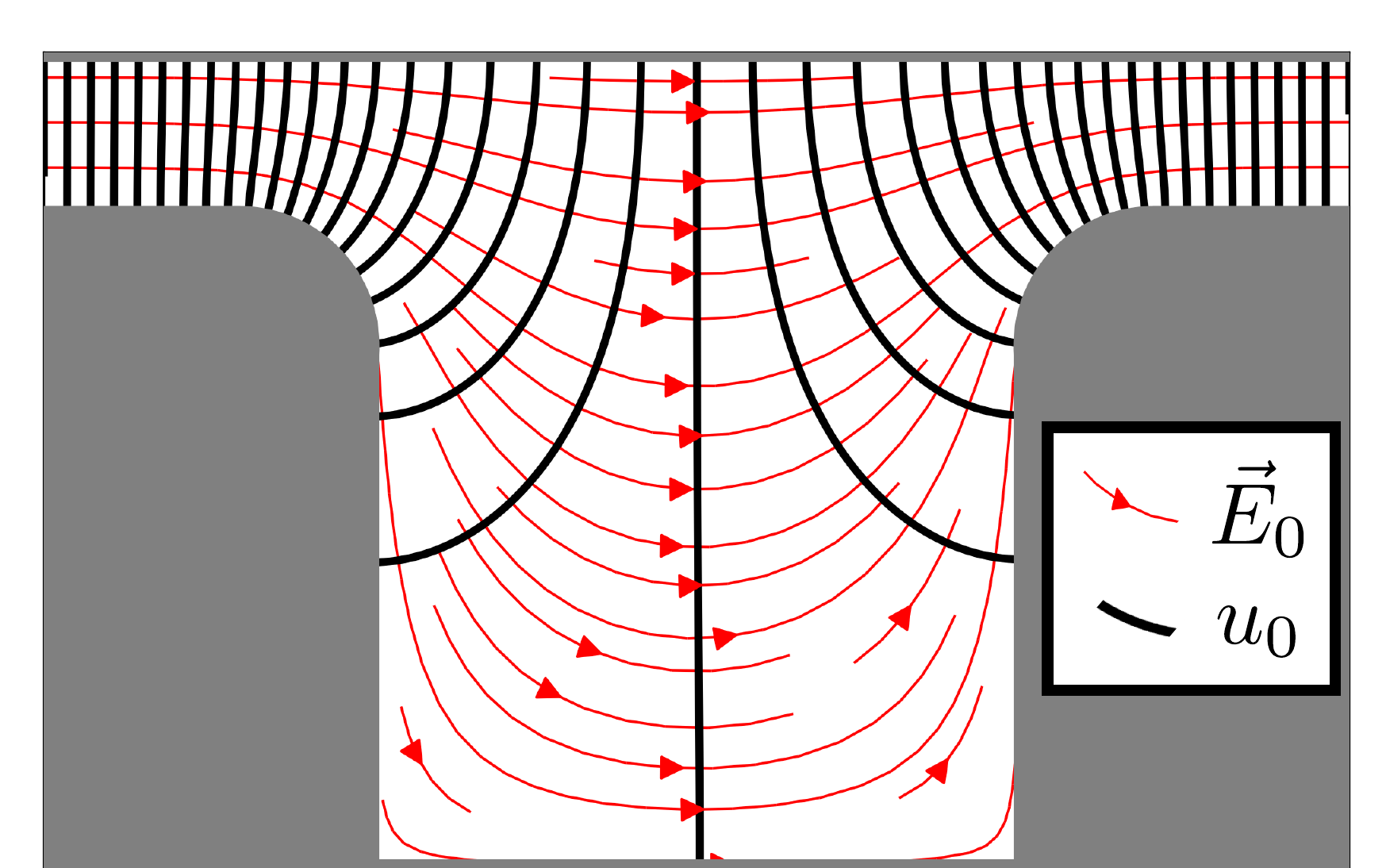}
\caption{Contour plots of the baseline electric potential $u_0$ (black) and field $\vec{E}_0$ (red) computed by the NNM.}
\label{fig:contour}
\end{figure}

The baseline electric field $\vec{E}_0$ used to drive particle motion in the slit-well device (Eqn.~\ref{eqn:reduced_g0}) was computed using the NNM, as described in \citet{Magill2020field}. 
That is, the baseline electric potential $u_0$ was solved using Laplace's equation over a voltage drop of two units from the left slit wall to the right slit wall.
The electric field was then computed using the relation $\vec{E}_0=\nabla u_0$.
The red and black contour lines in Fig.~\ref{fig:contour} correspond to the electric field $\vec{E}_0$ and electric potential $u_0$ respectively, inside the slit-well MNFD.

\FloatBarrier

\FloatBarrier

\section{RUNTIME COMPARISON \label{app:runtime}}

\FloatBarrier

The MFPT and effective mobility of nanoparticles traversing the slit-well MNFD were obtained using BD simulations, FEM and NNM.
Table~\ref{tab:runtimes} shows the runtime, in minutes, of each method used to solve the MFPT at fixed choices parameter values.
Four choices of the parameters are included, illustrating that runtimes were fairly independent of parameters for NNM and FEM, but depended strongly on parameters for BD.
Table~\ref{tab:runtimes-param} shows the runtime, in days, of each method used to solve the MFPT over large regions of parameter space.
As implemented, the various parameterized NNM methods all have comparable runtimes to one another.
Moreover, the total runtime of the 8099 FEM solutions exceeds the mean runtime for the parameterized NNM methods.

Optimizing runtime was not a goal of the current work.
The implementations of each of the algorithms studied here (NNM, BD, FEM) can undoubtedly be improved upon to substantially decrease the runtimes from those reported in Table~\ref{tab:runtimes} and \ref{tab:runtimes-param}.
Moreover, judicious use of parallelization across GPUs and/or CPUs, as applicable, could provide further improvements to each of the methods.
Thus, the runtimes included here are provided for reference only, and a more careful comparison is left to future work.

\begin{table}[ht]
	\begin{tabular}{@{} *{4}{c} @{}}
	\midrule
	\midrule
	\headercell{ \\ \hspace{10pt} Parameters $(\lambda, \sigma)$} \hspace{10pt} & \multicolumn{3}{c@{}}{Runtime (minutes)}\\
	\cmidrule(l){2-4}
	& NNM &  BD & FEM    \\ 
	\midrule
	  $(5.0, 0.125)$  & 136.95 &  5.02  & 2.25 \\
	  $(50.0, 0.125)$ & 138.37 &  1.03  & 2.15 \\
	  $(5.0, 0.625)$  & 126.57 &  12.80 & 2.18 \\
	  $(50.0, 0.625)$ & 133.63 &  1.12  & 2.13 \\
	\midrule
	\midrule
	\end{tabular}
\caption{Comparison of computational time (in minutes) of the NNM, BD, and FEM methods used to compute the MFPT at fixed parameter values.}
\label{tab:runtimes} 
\end{table}

\begin{table}[ht]
	\begin{tabular}{cc}
	\midrule
	\midrule
	Method &  Mean Runtime (days) \\ 
	\midrule	  
	  NNM parameterized by $\lambda$ & 6.33 \\
	  NNM parameterized by $\sigma$ & 7.79 \\
	  NNM parameterized by $(\lambda,\sigma)$ & 7.66 \\
	  High-resolution FEM sampling & 12.18 \\
	\midrule
	\midrule
	\end{tabular}
\caption{Comparison of computational time (in days) of the various methods used to compute the MFPT over ranges of parameter space.}
\label{tab:runtimes-param} 
\end{table}

\FloatBarrier

\setlength{\bibsep}{10pt}
\bibliographystyle{unsrtnat}
\bibliography{2020-07-mfptrefs}

\end{document}